\def\isabridged{1}
\def\isarxiv{1}
\ifdefined\isarxiv
\documentclass[a4paper, twocolumn, 10pt]{article}
\usepackage[top=2cm, bottom=2cm, left=1.75cm, right=1.75cm]{geometry}
\usepackage[hyphens]{url}
\usepackage{hyperref}
\date{}
\else
\documentclass[letterpaper,twocolumn,10pt]{article}
\PassOptionsToPackage{hyphens}{url}
\usepackage{usenix}

\fi
\usepackage[textsize=footnotesize]{todonotes}
\usepackage{graphicx}
\usepackage{xspace}
\usepackage{tikz}
\usepackage{glossaries}
\usepackage{pifont}
\usepackage{listings}
\usepackage{pifont}
\usepackage{algorithmic}
\usepackage{textcomp}
\usepackage[dvipsnames]{xcolor}
\usepackage{paralist}
\usepackage{enumitem}
\usepackage{hyperref}
\usepackage{cleveref}
\usepackage{float}
\usepackage{subcaption}
\usepackage{fvextra}
\usepackage[textsize=footnotesize]{todonotes}
\usepackage{glossaries}
\glsdisablehyper
\usepackage{caption}
\usepackage{subcaption}
\usepackage{listings}
\usepackage{xspace}
\usepackage{comment}
\usepackage{fancyvrb}
\usepackage{graphicx,txfonts}
\usepackage{multirow}
\usepackage{tikz}
\usepackage{multicol}
\usepackage{verbatim}
\usepackage{framed}
\usepackage{setspace}
\usetikzlibrary{positioning,arrows.meta,shapes.geometric}
\usetikzlibrary{
    backgrounds,
    fit,
    positioning
}
\usepackage{fontenc}
\usepackage{bytefield}
\usepackage{booktabs}
\usepackage[dvipsnames]{xcolor}
\usepackage{mwe}
\usepackage{wrapfig}
\usepackage{color, colortbl}
\usepackage{threeparttable}
\usepackage{array}
\usepackage{adjustbox}
\usepackage{wasysym}
\usepackage{xparse}
\usepackage[htt]{hyphenat}
\usepackage[hashEnumerators,smartEllipses]{markdown}
\usepackage{rotating}
\usepackage{todonotes}

\crefformat{section}{\S#2#1#3}
\crefformat{subsection}{\S#2#1#3}
\crefformat{subsubsection}{\S#2#1#3}
\newif\ifabridged
\newif\ifnotabridged
\newif\ifanonymous
\newif\ifnotanonymous
\newif\ifreviewer
\newif\ifnotreviewer
\newif\iftags
\newif\ifnottags

\ifdefined\isabridged
\abridgedtrue
\fi

\ifabridged\notabridgedfalse
\else\notabridgedtrue
\fi

\ifdefined\isanonymous
\anonymoustrue
\fi

\ifanonymous\notanonymousfalse
\else\notanonymoustrue
\fi

\newif\ifarxiv
\newif\ifnotarxiv

\ifdefined\isarxiv
\arxivtrue
\fi

\ifarxiv\notarxivfalse
\else\notarxivtrue
\fi

\ifdefined\isreviewer
\reviewertrue
\fi

\ifreviewer\notreviewerfalse
\else\notreviewertrue
\fi

\ifdefined\hashtags
\tagstrue
\fi

\iftags\nottagsfalse
\else\nottagstrue
\fi

\glsaddkey
    {hyphenated}
    {\relax}
    {\glsentryhyph}
    {\Glsentryhyph}
    {\glshyph}
    {\Glshyph}
    {\GLShyph}

\DeclareFontFamily{OT1}{mathc}{}
\DeclareFontShape{OT1}{mathc}{m}{it}{<-> mathc10}{}
\DeclareMathAlphabet{\mathabxcal}{OT1}{mathc}{m}{it}
\newacronym{abi}{ABI}{application binary interface}
\newacronym{ai}{AI}{artificial intelligence}
\newacronym{adv}{$\mathabxcal{Adv}$}{adversary}
\newacronym{alloc}{$\mathabxcal{Alloc}$}{allocator}
\newacronym{alu}{ALU}{arithmetic logic unit}
\newacronym{api}{API}{application programming interface}
\newacronym{asan}{ASan}{AddressSanitizer}
\newacronym{dma}{DMA}{direct memory access}
\newacronym[hyphenated={colored-capability},longplural={colored capabilities}]{cc}{CC}{colored capability}
\newacronym{ccsettype}{\texttt{ccsettype}}{colored capability set type instruction}
\newacronym{cisa}{CISA}{Cybersecurity & Infrastructure Security Agency}
\newacronym{cheri}{CHERI}{Capability Hardware Enhanced RISC Instructions}
\newacronym{clc}{\texttt{clc}}{load capability via capability}
\newacronym{comp}{$\mathabxcal{Comp}$}{compiler toolchain}
\newacronym{cpu}{CPU}{central processing unit}
\newacronym{crg}{\texttt{CRG}}{capability read generation }
\newacronym{csc}{\texttt{csc}}{store capability via capability}
\newacronym{csp}{\texttt{csp}}{stack pointer capability}
\newacronym{csr}{CSR}{control and status register}
\newacronym{cve}{CVE}{common vulnerability enumeration}
\newacronym{cwe}{CWE}{common weakness enumeration}
\newacronym{cw}{\texttt{CW}}{capability write}
\newacronym{df}{DF}{double-free}
\newacronym{eda}{EDA}{electronic design automation}
\newacronym{elf}{ELF}{executable and linkable format}
\newacronym{etsi}{ETSI}{European Telecommunication Standards Institute}
\newacronym{fpga}{FPGA}{field-programmable gate array}
\newacronym[hyphenated={geometric-mean}]{gm}{g.m.}{geometric mean}
\newacronym{id}{ID}{identifier}
\newacronym[longplural={instruction set architectures}]{isa}{ISA}{instruction-set architecture}
\newacronym{ir}{IR}{intermediate representation}
\newacronym{ip}{IP}{intellectual property}
\newacronym{jit}{JIT}{just-in-time}
\newacronym{lsu}{LSU}{load-store unit}
\newacronym{lsq}{LSQ}{Load/Store Queue}
\newacronym{gep}{\textsf{GEP}}{\textsf{GetElementPtr}}
\newacronym[hyphenated={provenance-ID},longplural={provenance identifiers}]{pid}{provenance ID}{provenance identifier}
\newacronym{nand}{NAND}{Not-AND}
\newacronym{ncsc}{NCSC}{National Cyber Security Centre}
\newacronym{nist}{NIST}{National Institute of Standards and Technology}
\newacronym{mte}{MTE}{Memory Tagging Extension}
\newacronym{msrc}{MSRC}{Microsoft Security Response Center}
\newacronym{mrs}{MRS}{\texttt{malloc} revocation shim}
\newacronym[hyphenated={operating-system}]{os}{OS}{operating system}
\newacronym{otype}{\texttt{otype}}{object type}
\newacronym{otth}{OTYPETH}{otype threshold}
\newacronym{u}{\texttt{U}}{user mode access allowed}
\newacronym{ucrg}{\texttt{UCRG}}{user capability read generation}
\newacronym{pa}{PA}{pointer authentication}
\newacronym{pc}{PC}{program counter}
\newacronym{pcc}{\texttt{pcc}}{program counter capability}
\newacronym{picasso}{\texttt{PICASSO}}{Provenance Indirection-enabled CHERI Architecture for Scarcely Swept Objects}
\newacronym{poc}{PoC}{proof-of-concept}
\newacronym{pte}{PTE}{page table entry}
\newacronym{ptlb}{PTLB}{provenance-translation lookaside buffer}
\newacronym{pvb}{PVB}{provenance-validity bit}
\newacronym[hyphenated={provenance-validity-table}]{pvt}{PVT}{provenance-validity table}
\newacronym{pvtr}{PVTR}{provenance-validity table register}
\newacronym{ross}{\texttt{ROSS}}{Revocation-Orchestrating System Service}
\newacronym{rtl}{RTL}{register-transfer level}
\newacronym{rtos}{RTOS}{real-time operating system}
\newacronym{rss}{RSS}{resident set size}
\newacronym{rv64y}{RV64Y}{64-bit RISC-V}
\newacronym{rof}{RoF}{revoke-on-free}
\newacronym{sard}{SARD}{Software Assurance Reference Dataset}
\newacronym{sp}{SP}{stack pointer}
\newacronym{soc}{SoC}{system-on-chip}
\newacronym{ssa}{SSA}{static single-assignment}
\newacronym{suar}{SUAR}{stack use-after-return}

\newacronym{tlb}{TLB}{translation lookaside buffer}
\newacronym{tls}{TLS}{thread local storage}
\newacronym{toctou}{TOCTOU}{time-of-check to time-of-use}
\newacronym{tps}{TPS}{transactions per second}
\newacronym{ts}{TS}{Technical Specification}
\newacronym{uaf}{UAF}{use-after-free}
\newacronym{uar}{UAR}{use-after-reallocation}
\newacronym{unr}{\texttt{unr}}{unit number}
\newacronym{vmem}{VM}{virtual memory}
\newacronym{wns}{WNS}{worst negative slack}
\newacronym{qps}{QPS}{Queries Per Second}
\newcommand*\circledblk[1]{\tikz[baseline=(char.base)]{
            \node[shape=circle,fill,inner sep=0pt] (char) {\textcolor{white}{#1}};}}

\renewcommand{\paragraph}{\noindent\textbf}

\newcommand{\dOne}{\ding{182}\xspace}
\newcommand{\dTwo}{\ding{183}\xspace}
\newcommand{\dThree}{\ding{184}\xspace}
\newcommand{\dFour}{\ding{185}\xspace}
\newcommand{\dFive}{\ding{186}\xspace}
\newcommand{\dSix}{\ding{187}\xspace}
\newcommand{\dSeven}{\ding{188}\xspace}
\newcommand{\dEight}{\ding{189}\xspace}
\newcommand{\dNine}{\ding{190}\xspace}
\newcommand{\dTen}{\ding{191}\xspace}
\newcommand{\dEleven}{\circledblk{\scriptsize11}\xspace}
\newcommand{\dTwelve}{\circledblk{\scriptsize12}\xspace}
\newcommand{\dCOne}{\ding{192}\xspace}
\newcommand{\dCTwo}{\ding{193}\xspace}
\newcommand{\dCThree}{\ding{194}\xspace}
\newcommand{\dCFour}{\ding{195}\xspace}
\newcommand{\dCFive}{\ding{196}\xspace}
\newcommand{\dCSix}{\ding{197}\xspace}
\newcommand{\dCSeven}{\ding{198}\xspace}

\newcommand{\dCNine}{\ding{200}\xspace}

\newlength{\dingwidth}
\newcounter{tncnt}

\newcounter{mgcnt}

\newcounter{jtcnt}

\newcounter{hecnt}

\newcounter{rscnt}
\newcommand{\new}[1]{#1}

\newcommand{\Adv}{\gls{adv}\xspace}
\newcommand{\Alloc}{\gls{alloc}\xspace}
\newcommand{\Comp}{\gls{comp}\xspace}
\newcommand{\PICASSO}{\acrshort{picasso}\xspace}

\newcommand{\cc}{\glsentrydesc{cc}\xspace}
\newcommand{\ccs}{\glsentrylongpl{cc}\xspace}

\newcommand{\colorbitbox}[3]{
  \sbox0{\bitbox{#2}{#3}}
  \makebox[0pt][l]{\textcolor{#1}{\rule[-\dp0]{\wd0}{\ht0}}}
  \bitbox{#2}{#3}
}

\newcommand{\ccinactive}{invalid\xspace}

\newcommand{\ccpid}{\gls{pid}\xspace}
\newcommand{\ccpidhyph}{\glshyph{pid}\xspace}
\newcommand{\ccpids}{\glspl{pid}\xspace}

\newcommand{\ccpidslong}{\glsentrylongpl{pid}\xspace}

\newcommand{\FRESCO}{\texttt{FRESCO}\xspace}
\newcommand{\libfresco}{\texttt{libfresco}\xspace}

\newcommand{\stackcoloring}{stack coloring\xspace}
\newcommand{\Stackcoloring}{Stack coloring\xspace}
\newcommand{\StackColoring}{Stack Coloring\xspace}

\newcommand{\colorsegmentationlong}{capability-color segmentation\xspace}
\newcommand{\Colorsegmentationlong}{Capability-color segmentation\xspace}

\newcommand{\colorsegmentation}{color segmentation\xspace}
\newcommand{\Colorsegmentation}{Color segmentation\xspace}

\newcommand{\ColorSaver}{Color Saver\xspace}
\newcommand{\initialclear}{initial clear\xspace}
\newcommand{\initialclears}{initial clears\xspace}

\newcommand{\anonURLClicky}{\href{https://anonymous.4open.science/r/2485414636761283591208227}{this link}\xspace}

\lstdefinelanguage{riscv}{
  morekeywords={addi,addiw,lui,ld,sd,lw,sw,ret,call,mv,li,blt,beq,bne,bnez,beqz,bge,slli,j,jr},
  morekeywords=[2]{ccsettype,cgetype,cincoffset,csetbounds,clc,csc,cld,csd,csw,clw,cret,cmove,ccall},
  keywordstyle=\color{blue}\bfseries,
  keywordstyle=[2]\color{teal}\bfseries,
  comment=[l]{\#},
  commentstyle=\color{gray}\itshape,
  morestring=[b]",
  sensitive=false,
}

\makeatletter
\newcommand{\myfnsymbol}[1]{%
  \expandafter\@myfnsymbol\csname c@#1\endcsname
}
\newcommand{\@myfnsymbol}[1]{%
  \ifcase #1
  \or \TextOrMath{\textasteriskcentered}{*}%
  \or \TextOrMath{\textdagger}{\dagger}%
  \or \TextOrMath{$\ddagger$}{\ddagger}%
  \or \TextOrMath{\textsection}{\S}%
  \or \TextOrMath{\textbardbl}{\|}%
  \or \TextOrMath{\P}{\P}%
  \or \TextOrMath{\#}{\#}%
  \or \TextOrMath{$\Delta$}{\Delta}%
  \fi
}
\newcommand{\equalContribution}{\@myfnsymbol{1}}
\newcommand{\affiliationA}{\@myfnsymbol{2}}
\newcommand{\affiliationB}{\@myfnsymbol{3}}
\newcommand{\affiliationC}{\@myfnsymbol{4}}
\newcommand{\affiliationD}{\@myfnsymbol{5}}
\newcommand{\affiliationE}{\@myfnsymbol{6}}
\newcommand{\affiliationF}{\@myfnsymbol{7}}
\newcommand{\affiliationG}{\@myfnsymbol{8}}
\newcommand{\affiliationH}{\@myfnsymbol{9}}
\makeatother

\title{\FRESCO: Complete and Scalable Temporal Safety for \\ CHERI Application Processors}

\ifnotanonymous
\author{Merve G\"{u}lmez\thanks{The authors contributed equally to this research.}\textsuperscript{\,\,,\affiliationA}, Nils Jordan\textsuperscript{\equalContribution,\affiliationA}, 
Jialun Zhang\textsuperscript{\equalContribution,\affiliationB},
Hossam ElAtali\textsuperscript{\affiliationC},\\
Gang Tan\textsuperscript{\affiliationB},
N. Asokan\textsuperscript{\affiliationC,\affiliationD}, Thomas Nyman\textsuperscript{\affiliationE}\\
\textit{\textsuperscript{\affiliationA}Ericsson Security Research, \textsuperscript{\affiliationB}Pennsylvania State University, \textsuperscript{\affiliationC}University of Waterloo,}\\\textit{ \textsuperscript{\affiliationD}KTH Royal Institute of Technology, \textsuperscript{\affiliationE}Ericsson Product Security} \\
\{merve.gulmez,nils.jordan,thomas.nyman\}@ericsson.com, \{jialun.zhang,gtan\}@psu.edu\\hossam.elatali@uwaterloo.ca, asokan@acm.org}
\fi

\begin{document}

\maketitle

\begin{abstract}
    
\ifarxiv
\small
\fi

\acrshort{cheri} provides hardware-enforced spatial memory safety. While prior work extends it with heap temporal safety, \emph{stack use-after-return} remains unaddressed.
Existing defenses fall short: compiler analysis reliably catches only references that escape as function return values, while dynamic sanitizers impose overheads that preclude production deployment.

We present \FRESCO, built on the principle that a stack capability must not outlive the frame that created it.
\FRESCO ``\emph{colors}'' the stack pointer with per-invocation \ccpidslong; every capability derived from it inherits that lifetime and is hardware-invalidated the moment the function exits, regardless of how or where it escaped.
Because stack frames retire orders of magnitude more frequently than heap allocations, \FRESCO manages the resulting color pressure through:
\begin{inparaenum}[1)]
    \item \emph{\ColorSaver}, a static capability-aware escape analysis that confines coloring to functions needing it, and whose core algorithm we mechanically verify in Rocq, and
    \item \emph{\colorsegmentationlong}, which partitions memory into disjoint segments, each with an independent color namespace.
\end{inparaenum}
\Colorsegmentation lets stack and heap temporal safety coexist on one system, making \FRESCO the first hardware/software co-design to provide \textbf{complete and scalable temporal safety} for \acrshort{cheri} application processors.
We realize \FRESCO on the CHERI-RISC-V QEMU full-system emulator and the out-of-order CHERI-Toooba \acrshort{fpga} softcore, with software support in the \acrshort{cheri}-enabled Clang/LLVM compiler and CheriBSD \acrshort{os}. \FRESCO systematically prevents use-after-return, use-after-free, and double-free across the NIST Juliet Test Suite and 
\acrshort{cve}s, with only a small run-time overhead in SPEC \acrshort{cpu} ($\approx4\%$ g.m.), SQLite, and PostgreSQL ($\approx10-14\%$).
 \end{abstract}

\glsresetall
\section{Introduction}

\gls{cheri} is a prominent design for hardware-assisted memory safety. By using capabilities---hardware-supported descriptions of permissions---instead of conventional pointers, it inherently provides hardware-enforced spatial safety for C and C++. Prior work augments CHERI to also provide temporal-safety guarantees by mitigating heap-based \glsdesc{uaf}/\glsdesc{uar} vulnerabilities~\cite{WesleyFilardo20,Filardo24,Gulmez26,Wang26}.
However, they only cover heap allocation in \gls{cheri} application processors, leaving stack \emph{use-after-return}  unmitigated. Stack use-after-return bugs occur when a stack-allocated object is used after the function where the object is defined has returned.
Compiler-based escape analysis can reliably detect use-after-return when stack-local references escape the allocating function as return values (as returning such a reference is always an error), but escapes via, e.g., pointers embedded into heap-allocated or global objects require dynamic analysis, e.g., sanitizers~\cite{LLVMTeam26,GCCTeam26}.

\noindent\textbf{This paper and contributions.}
Building on recent work introducing colored capabilities~\cite{Gulmez26} to \gls{cheri}, we design \emph{\stackcoloring}, a novel stack-focused temporal-safety scheme which deterministically protects against use-after-return. \Stackcoloring leverages the ability to simultaneously retract sets of colored capabilities at once without the need to fully revoke them. It turns the stack pointer capability into a colored capability ensuring that any capabilities that reference allocations in a particular stack frame reflect a stack allocation's \emph{provenance} through a `color' (provenance \acrlong{id}) assigned to precisely that frame. By retracting dangling capabilities to retired stack frames when a function returns, \stackcoloring prevents use-after-return conditions regardless of how references escape the functions their allocations belong to.

Since stack frames are retired orders of magnitude more frequently than heap allocations, we must scale colored capabilities well beyond the $\approx2M$ unique provenance \glspl{id} in prior work~\cite{Gulmez26} to support \stackcoloring.
We manage the resulting color pressure through two complementary mechanisms:
\begin{inparaenum}[1)]
\item \emph{\ColorSaver}: a compiler extension which confines coloring to functions that need it through a capability-aware escape analysis and prevents stale capability reads from reused stack memory; we mechanically verify the soundness of its escape classifier in Rocq~\cite{Herbelin26}, and
\item \emph{\colorsegmentationlong}: a hardware feature that associates segments of memory with distinct \glspl{pvt}, allowing $\approx2M$ colors \emph{per memory segment}.
\end{inparaenum}

\Colorsegmentation allows stack and heap temporal-safety enforcement leveraging colored capabilities to \emph{coexist} on the same system.
Further, we demonstrate how \colorsegmentation improves performance of heap temporal-safety enforcement in \emph{color-segmented allocators}.
This enables us to realize a full temporal-safety solution for \gls{cheri} in \FRESCO, the first hardware/software co-design providing complete and scalable temporal safety for \gls{cheri} application processors.
In summary, our contributions are:
\begin{itemize}[nosep]
    \item A \textbf{complete temporal-safety solution for \gls{cheri}} combining \emph{\stackcoloring}, a novel stack temporal-safety scheme leveraging colored capabilities~(\Cref{sec:stackcoloring}); and a
    \emph{color-segmented allocator} enforcing temporal safety of heap allocations~(\Cref{sec:segmentedalloc-design}). We implement \stackcoloring in \gls{cheri}-enabled Clang/LLVM~(\Cref{sec:instrumentation}) and a jemalloc-based color-segmented allocator in CheriBSD~(\Cref{sec:segmentedalloc}).
    \item \emph{\ColorSaver}, combining capability-aware escape analysis and stale-read prevention to reduce color use without weakening guarantees (\Cref{sec:color-saver}). We integrate \ColorSaver with \gls{cheri}-enabled Clang/LLVM (\Cref{sec:color-saver-implementation}) and mechanically verify the basic escape classifier in Rocq (\Cref{subsec:escape-analysis}).
    \item \emph{\Colorsegmentationlong}, an extension to colored capabilities~\cite{Gulmez26} scaling them well beyond $\approx2M$ provenance \glspl{id} (\Cref{sec:colorsegmentation}). We implement \colorsegmentation in \FRESCO, a \acrshort{cpu} design based on CHERI-RISC-V realized in the QEMU full-system emulator and CHERI-Toooba~\cite{CTSRD24a} \acrshort{fpga} softcore (\Cref{sec:implementation}).
    \item Showing \FRESCO systematically prevents use-after-return, use-after-free, and double-free conditions in \acrshort{nist} Juliet test cases and real-world
    \acrshort{cve}s (\Cref{sec:seceval}).
    
    \item Demonstrating the efficacy of the full solution running on \FRESCO in the SPEC \acrshort{cpu} benchmarks and real-world SQLite, PostgreSQL, and gRPC benchmarks (\Cref{sec:perfeval}).
\end{itemize}
\smallskip
\ifnotarxiv
We make the full source code and \acrshort{rtl}-implementation of \FRESCO available for review at \anonURLClicky and plan to open-source the implementation artifacts upon acceptance. 
\fi

\section{Background}

\subsection{CHERI}\label{sec:cheri}
\ifnotabridged
\begin{figure}[t]
    \begin{bytefield}[bitwidth=0.3em]{1}
        \begin{rightwordgroup}{\dCOne{} 1-bit Validity tag}
            \colorbitbox{black}{1}{}
        \end{rightwordgroup} \\
        \bitbox[]{1}{} 
    \end{bytefield}\\
    \begin{bytefield}[bitwidth=0.37em]{64}
        \bitbox{17}{\dCTwo{} Permissions} & \colorbitbox{lightgray}{2}{} &
        \bitbox{18}{\dCThree{} Object type} & \bitbox{27}{\dCFour{} Bounds} \\
        \bitbox{64}{\dCFive{} Baseline architecture address} \\
    \end{bytefield}
    \caption{In-memory representation of \gls{cheri} capabilities adapted from Watson et al.~\cite{Watson19}}\label{fig:chericap}
\end{figure}
\fi

\gls{cheri} is an \gls{isa} extension that augments conventional \glspl{isa} with a \emph{capability-based} hardware-software co-design for memory protection.
Hardware-supported capabilities enforce \emph{spatial safety} for code or data pointers.
The \gls{cheri} \gls{isa} specification~\cite{Watson23a} defines capability representations in registers and memory, along with instructions to manipulate them safely. 

\ifnotabridged As shown in \Cref{fig:chericap}, a\else A \fi \gls{cheri} capability is twice the width of a native pointer type---128~bits on 64-bit platforms---and includes one additional \emph{validity-tag} bit\ifnotabridged \dCOne\fi, stored separately to prevent tampering by  non-capability-aware instructions.
These tags are preserved by valid, capability-aware instructions but are invalidated by unauthorized manipulation or injection of arbitrary capabilities.
In \gls{cheri} \glspl{isa}, general-purpose and address registers are extended to hold full capabilities.
CHERI-RISC-V stores the \gls{pc} and \gls{sp} in the \gls{pcc} and \gls{csp} registers, respectively.
Each capability includes:
\begin{itemize}[leftmargin=*, nosep]
    \item \textbf{Permissions}\ifnotabridged \dCTwo\fi: A bitmask defining allowed operations.
    \item \textbf{Object type} (\acrshort{otype})\ifnotabridged \dCThree\fi: A signed integer enabling temporary ``sealing'' and ``unsealing'' of capabilities for opaque pointer and fine-grained in-process isolation.
    \item \textbf{Bounds}\ifnotabridged \dCFour\fi: The valid memory range relative to \ifnotabridged the baseline architecture\fi address\ifnotabridged \dCFive, using a compressed encoding~\cite{Woodruff19} to reduce the space taken by a capability at the cost of stricter alignment for larger object allocations\fi.
\end{itemize}

\ifnotabridged
\paragraph{CHERI RISC-V Calling Convention.}
The CHERI RISC-V \emph{pure-capability} calling convention adapts the standard RISC-V \gls{abi} by replacing traditional integer register mappings (\texttt{x0}--\texttt{x31}) with capability registers (\texttt{c0}--\texttt{c31}).
During the subroutine prologue, after the stack pointer capability (\texttt{csp}/\texttt{c2}) is decremented to allocate space for the stack frame, any callee-saved capability registers intended for use by the function are immediately preserved (spilled) by storing their original values to the stack.
Following these spills, the frame pointer capability (\texttt{cs0}/\texttt{c8}) is updated by copying the incremented \texttt{csp}. Subsequently, local variables are allocated and initialized within the frame, relative to \texttt{cs0}, before the subroutine body begins execution.

\paragraph{Inherent \gls{cheri} security properties.}\fi
\gls{cheri} enforces spatial safety by assigning each allocation a capability describing its valid range and permissions.
New capabilities are derived from existing ones, while maintaining \emph{monotonicity} which ensures that new capabilities cannot exceed the permissions or bounds of their parent.
Sealed capabilities for compartmentalization and exception handling introduce limited non-monotonicity.
Extensions to \gls{cheri} have also explored sandboxing~\cite{Chisnall17}, initialization safety~\cite{Georges21,Gulmez25a}, safe speculation~\cite{Fuchs23, Fuchs24}, and side-channel resistance~\cite{ElAtali25}.

\subsection{CHERI Temporal Safety}\label{sec:cheri-temporal-safety}

\paragraph{Cornucopia \& Cornucopia Reloaded.}
Cornucopia~\cite{WesleyFilardo20} provides deterministic \gls{uar} detection for heap allocations on \gls{cheri}: freed memory is quarantined and marked in a shadow bitmap, once quarantine exceeds a threshold, a kernel-performed revocation sweep clears the validity tags of stale capabilities before memory is reused.
Quarantine leaves dangling pointers valid, obscuring \gls{uaf} defects, and inflates memory overhead due to delayed reuse. The threshold trades mitigation completeness for performance, as \emph{revoke-on-free} is prohibitively expensive.
Even with concurrent sweeping, a stop-the-world phase is required to scan register files, recently modified pages and kernel-held capabilities.
Cornucopia Reloaded~\cite{Filardo24} removes global pauses by using capability \emph{load barriers}, which trap capability loads from pages not yet swept, but quarantine remains: memory freed during a sweep epoch cannot be reused until the next sweep completes.
CheriBSD includes Cornucopia Reloaded via the \gls{mrs}, which wraps BSD libc allocation functions (\texttt{malloc()} etc. and \texttt{free()}).

\paragraph{Colored Capabilities.}
\PICASSO\cite{Gulmez26} addresses the same problem, but eliminates the quarantine buffer and reduces frequency of sweeping revocation.
Its \emph{\ccs} introduce a deliberately limited form of indirection into \gls{cheri}'s otherwise indirection-free model,
decoupling the \textbf{validity of a capability's provenance} (the allocation it represents) from the \textbf{validity of the capability itself}.
Each colored capability carries its \emph{"color"} as a \ccpid in the \texttt{otype} field.
A corresponding \gls{pvb} is held in a separate, in-memory \gls{pvt}, indexed by the \ccpid, with in-use identifiers tracked by the \gls{unr} allocator (\ifnotabridged\Cref{sec:unralloc}\else Appendix~\ref{appx:unralloc}\fi).
Unlike memory-versioning schemes (\Cref{sec:relatedwork}), only the capability, not the backing memory, is colored.
Hardware consults the \gls{pvb} on every load or store issued through a colored capability, faulting once the \gls{pvb} is invalidated.
This \emph{retracts} all capabilities sharing provenance at once, disabling dereferencing dangling \ccs without permanently revoking them and closing the \acrshort{uaf}/\gls{uar} gap.
\PICASSO integrates into the CheriBSD \gls{mrs}, as a drop-in replacement for the default scheme.

To keep provenance checks off the critical \gls{cpu} path, \PICASSO caches recently accessed 128-bit \gls{pvt} words in a \gls{pvt}-buffer, a small (64-word, 4-way set-associative) cache structure indexed by the virtual \gls{pvb} address.
A colored load or store can read a cached \gls{pvb} without paying for \gls{ptlb} translation and \gls{pvt} load. Coherence is maintained by flushing the entire buffer on any fence instruction; correctness therefore relies on the \gls{mrs} issuing a fence after it toggles a \gls{pvt}, and on fences after page-table updates.

\ifnotabridged
\subsection{\PICASSO \acrshort{unr} allocator}\label{sec:unralloc}

The \gls{unr} allocator is PICASSO's adaptation of FreeBSD's BSD unit-number allocator for issuing and reclaiming \ccpids (colors).
It exhibits several scalability limitations rooted in its compressed representation of allocation state as a doubly-linked list of \emph{runs} and \emph{bitmaps}.
As colors become fragmented, allocation degrades from a constant-time run adjustment to a linear scan of bitmaps for the first clear bit, and reclaiming a single \gls{id} situated within a run forces that run to be split into three parts followed by a \emph{collapse-and-merge} maintenance step (see Appendix \Cref{appx:unralloc}).
Most significantly, FreeBSD's allocator lacks bulk reclamation, which PICASSO requires during revocation; calling the per-\gls{id} free function in a loop is prohibitively inefficient, so PICASSO instead introduces \texttt{free\_many\_unr}, which rebuilds the entire structure from scratch by allocating a new header, reconstructing state from the revoker's \gls{pvt} snapshot, and freeing the old header.
This rebuild incurs substantial memory overhead, since the old and new structures coexist during execution, and substantial run-time overhead from numerous internal node allocations, with further structural maintenance deferred to subsequent allocation calls. These costs are inherent to the design: mid-run frees necessitate restructuring, and scattered bulk frees require either individually modifying every affected node or discarding and rebuilding the whole structure, rendering the \gls{unr} allocator's data structures unsuited for in-place bulk updates.
\fi

\subsection{Remaining \gls{cheri} Temporal-Safety Gaps}\label{sec:temporal-safety-gaps}
State of the art temporal-safety schemes for \gls{cheri} still exhibit gaps in their effectiveness and scalability:

\paragraph{1) No stack temporal-safety enforcement.}
\PICASSO~\cite{Gulmez26}, like earlier CHERI work~\cite{WesleyFilardo20,Filardo24}, leaves use-after-return out of scope, on the basis that it is rare and ``unlikely to be exploitable''~\cite{Lee15}.
In practice, however, temporal-safety violations on stack-allocated objects  continue to yield medium- to high-severity disclosures to the \gls{cve} program~\cite{NVD26,NVD26a,NVD26b} (\Cref{sec:relatedwork}).
Proponents of memory-safety standardization have called on regulators to incentivize deployment of \emph{complete}, deterministic memory-safety technologies in the coming decade~\cite{Watson25} (\Cref{sec:discussion}), and advances in automating exploit development with \gls{ai} agents~\cite{Wang26a,Wang26b} add urgency to complete enforcement, including of stack use-after-return.

\paragraph{2) Bounded color space.}
\PICASSO's 21-bit identifiers admit just over two million concurrently distinguishable allocations per process:
that far exceeds the 16 afforded by 4-bit memory-tagging schemes~\cite{Filardo24,Filardo24a} (\Cref{sec:relatedwork}) and makes sweeps rare, but the pool remains finite, placing an upper limit on the number of concurrent live allocations the scheme can scale to. 
Enlarging the space does not scale freely; it consumes scarce capability bits (competing with other \gls{cheri} extensions), inflates the \gls{pvt} and the cost of used-identifier bookkeeping.
\ifnotabridged Gülmez et al.~\cite{Gulmez26} note that a pointer-masking design supporting $2^{37}$ identifiers would demand an impractical 16 GiB \gls{pvt}.\fi

\paragraph{3) Residual overhead on allocation-heavy workloads.}
Both designs in \Cref{sec:cheri-temporal-safety} pay a cost proportional to the number of revocations, but differ in what triggers them: Cornucopia's quarantine buffer threshold, versus exhaustion of \PICASSO's global color space: omnetpp, the SPEC CPU2006 INT benchmark allocating enough to trigger revocation in \PICASSO, costs 38\% in run-time and roughly 80\% in memory~\cite{Gulmez26}.
This is an obstacle to deployment in allocation-intensive programs.

\section{Problem Statement}\label{sec:problem-statement}

\subsection{Goals}\label{sec:goals-and-challenges}

Our overall objective is to bridge the gaps 1) -- 3) discussed in \Cref{sec:temporal-safety-gaps} by achieving the following goals:
\begin{enumerate}[label=\textbf{G\arabic*},labelsep*=1em,itemsep=2pt]
  \item\label{goal:full} \textbf{Full temporal safety.} Deterministically protect against \emph{both} use-after-free and use-after-return of heap and stack allocations, closing the stack temporal-safety gap in prior \gls{cheri} temporal-safety designs.
  \item\label{goal:nocolorceiling} \textbf{Scalability without a color ceiling.} The number of simultaneously live objects managed by \ccs must be bounded by available memory rather than by the bits the capability format reserves for the \ccpid. Long-running and allocation-heavy workloads must remain deployable without widening that field.
  \item\label{goal:boundedmetadata} \textbf{Bounded revocation cost}. Revocation must be infrequent. Its trigger must not be exhaustion of a single global identifier pool, and each cycle's cost must be proportional to the live state it reclaims rather than by cumulative allocation history or by fragmentation of the identifier space.
\end{enumerate}

\subsection{System and Adversary Model}\label{sec:system-and-adversary-model}

\paragraph{System Model.}
We assume the same system model as \gls{cheri}~\cite{WesleyFilardo20} and  \PICASSO~\cite{Gulmez26} and that \gls{cheri}'s spatial-safety properties hold for both heap- and stack-allocated objects.
We rely on \Comp and heap \Alloc satisfying the following assumptions:

\begin{enumerate}[label=\textbf{A\arabic*},labelsep*=0pt,itemsep=2pt]
   \item:~\Comp is trusted and correctly instruments function prologues and epilogues that manage stack-frames.\label{asm:comp-is-trusted}
    \item:~\Alloc is part of the trusted runtime.\label{asm:alloc-is-trusted}
    \item:~\Comp derives capabilities to stack-allocated objects with bounds no wider than the enclosing stack frame.\label{asm:correct-stack-bounds}
    \item:~\Alloc maintains correct bounds on capabilities it issues.\label{asm:correct-heap-bounds} 
    \item:~\Alloc can discern malformed input to \texttt{free()}.\label{asm:malformed-free}
 
\end{enumerate}

\paragraph{Adversary Model.}
We assume an \Adv who manipulates program behavior through malicious input.
Exploiting defects in the victim program, such as faulty input validation, \Adv can influence it to 
\begin{inparaenum}[i)]
    \item invoke arbitrary sequences of \texttt{malloc()} and \texttt{free()} calls (or C++ \texttt{new} and \texttt{delete}), and
    \item trigger arbitrary sequences of function calls and returns, thereby controlling the allocation, deallocation, and reuse of stack frames.  
\end{inparaenum}
In particular, \Adv can exploit program defects in which a capability to a stack-allocated object escapes its defining frame---by being returned, stored to heap- or statically-allocated memory, leaked from a retired stack frame through an initialization-safety violation, or passed to a concurrent thread---and is dereferenced after the frame is deallocated (use-after-return, CWE-562~\cite{MITRE24}), including after the frame's memory has been reused by a subsequent call.

\gls{cheri}'s architectural guarantees prevent \Adv from forging capabilities.
\Adv cannot influence compiler-emitted stack management beyond directing control flow through legitimate calls and returns (\ref{asm:comp-is-trusted}), and can interact with \Alloc only through \texttt{malloc()} and \texttt{free()}, not by tampering with its data structures, control-data, or code (\ref{asm:alloc-is-trusted}).
This is consistent with the adversary model of prior work~\cite{WesleyFilardo20, Filardo24, Gulmez26}.

\gls{cheri} capabilities are monotonically nonincreasing; software can narrow, but never widen, a derived capability's permissions and bounds relative to its parent.
Consequently, \ref{asm:correct-stack-bounds} implies that stack-directed capabilities lie within their (once-) enclosing frames, and \ref{asm:correct-heap-bounds} the analogous property for heap-directed capabilities and their (once) heap-allocated objects.
\ref{asm:malformed-free} ensures that \Alloc rejects capabilities not pointing to the start of a legitimate heap allocation;
\gls{mrs} rejects invalid capabilities and capabilities to non-heap-allocated data.

Uninitialized reads of stale stack- or heap-allocated capabilities are in scope.
Beyond that, uninitialized reads of memory allocations are out of scope, but can be systematically prevented by previously proposed \gls{cheri} extensions~\cite{Gulmez25a} or compiler-based uninitialized memory sanitization~\cite{Chow05,Milburn17,Joly20}.

Our completeness claim is consequently one of \emph{escape routes}, not of \emph{allocation granularity}.
Provenance is tracked per allocation on the heap and per function activation on the stack; within that granularity, protection is deterministic and independent of how references escape: whether returned in a register, stored into heap- or statically-allocated memory, recovered from reused stack memory, or passed to concurrent threads. 
We discuss intra-frame lifetime violations in \Cref{sec:discussion}. Consistent with prior \gls{cheri} implementations on application-class processors, we exclude transient-execution and other micro-architectural attacks and side-channels~\cite{Watson23b}.
We note, however, that similarly to \PICASSO~\cite{Gulmez26}, \FRESCO's provenance-validity checks are enforced before instruction commit: loads commit and stores issue only after the check succeeds, so temporal-safety violations never become architecturally visible; only transient covert channels remain out of scope.

\ifnotarxiv
\begin{figure}[t]
\begin{centering}
    \includegraphics[width=\columnwidth]{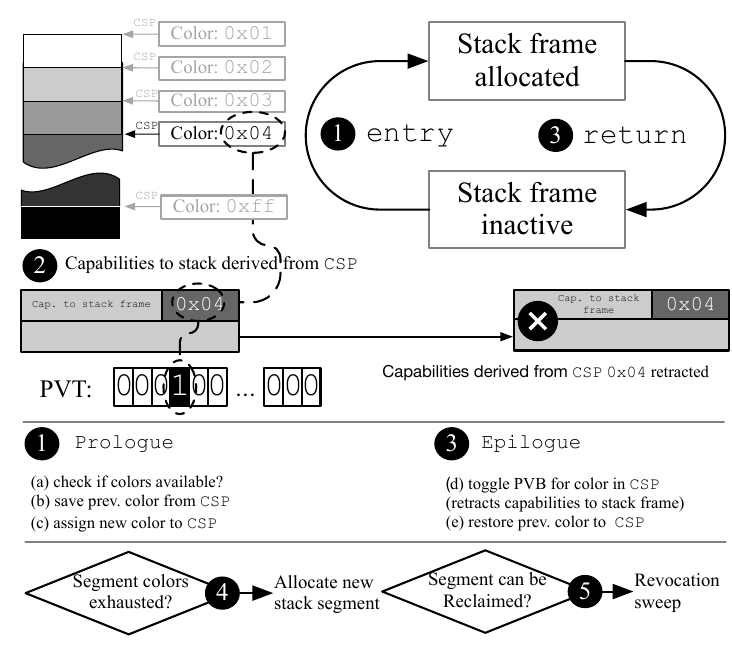}
    \caption{High-level overview of stack coloring.}\label{fig:stackcoloring}
\end{centering}
\end{figure}
\fi

\section{System Design}\label{sec:design}

\footnotetext[1]{Fresco is a mural technique executed on freshly laid, still-wet lime plaster. Analogously, \FRESCO's stack coloring ``paints'' the \gls{csp} in the prologue before the frame's ``plaster sets'' and the function body properly begins.}

\FRESCO achieves \ref{goal:full} by composing two \gls{cheri}-based temporal-safety mechanisms: \emph{allocation-provenance tracking} for the heap inspired by \PICASSO~\cite{Gulmez26} and summarized in \Cref{sec:cheri-temporal-safety}, and \emph{\stackcoloring}, our novel counterpart for stack allocations, which extends \ccs to track \emph{stack-frame provenance}.
\Cref{sec:stackcoloring} presents \stackcoloring, \ColorSaver, and \colorsegmentation; \Cref{sec:segmentedalloc-design} contrasts \FRESCO's and \PICASSO's approaches to heap temporal-safety.

\subsection{\StackColoring}\label{sec:stackcoloring}

\Cref{fig:stackcoloring} illustrates \stackcoloring: it turns the \gls{csp} into a \cc, enabling stack-frame provenance tracking, which associates each \cc with the color (\ccpid) of the frame holding the allocation it points to.
On function entry \dOne, the \gls{csp} receives a new color from a finite set, and \ccs subsequently derived from it inherit that color \dTwo.
The \gls{csp} retains its color for the duration of the function, so all capabilities to allocations within that frame share one color.
For simplicity, we refer to this as the \emph{stack frame's color} and depict frames as differently colored in \Cref{fig:stackcoloring}.
Strictly speaking, only \emph{\ccs} carry the color through their assigned \ccpid.

\ifarxiv
\begin{figure}[t]
\begin{centering}
    \includegraphics[width=\columnwidth]{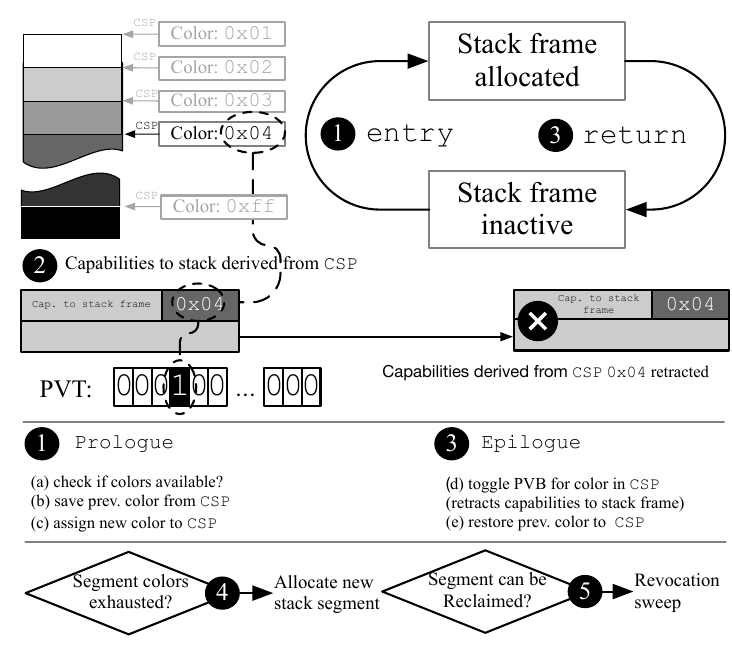}
    \caption{High-level overview of stack coloring.}\label{fig:stackcoloring}
\end{centering}
\end{figure}
\fi

When the function returns \dThree its stack frame becomes inactive, and de-referencing a capability into it would constitute a use-after-return.
\Stackcoloring prevents this by toggling the \gls{pvb} for the frame's color to \emph{\ccinactive}, thereby retracting every dangling pointer whose provenance ties it to the inactive frame.
Any subsequent dereference through an escaped capability bearing that color traps (\ref{goal:full}).

\paragraph{Frame coloring.}
\Stackcoloring assigns the \gls{csp} a unique color in the instrumented function prologue\footnotemark and retires that color in the epilogue.
Both are emitted by \gls{comp} (\Cref{sec:system-and-adversary-model}) which alone is entrusted with the \gls{ccsettype} that manipulates the \gls{csp}'s color.
Capabilities derived from the colored \gls{csp} inherit its   \gls{otype} and hence the frame's color.
Recall from \Cref{sec:cheri} and \Cref{sec:system-and-adversary-model} that untrusted software, lacking access to \gls{ccsettype}, cannot alter a capability's \gls{otype}.
On return, the epilogue toggles the \gls{pvb} for the retiring frame's color, restores the caller's color to \gls{csp}, and retires the old color from the assignable set.

\begin{figure*}[t]
\begin{centering}
    \includegraphics[width=2\columnwidth]{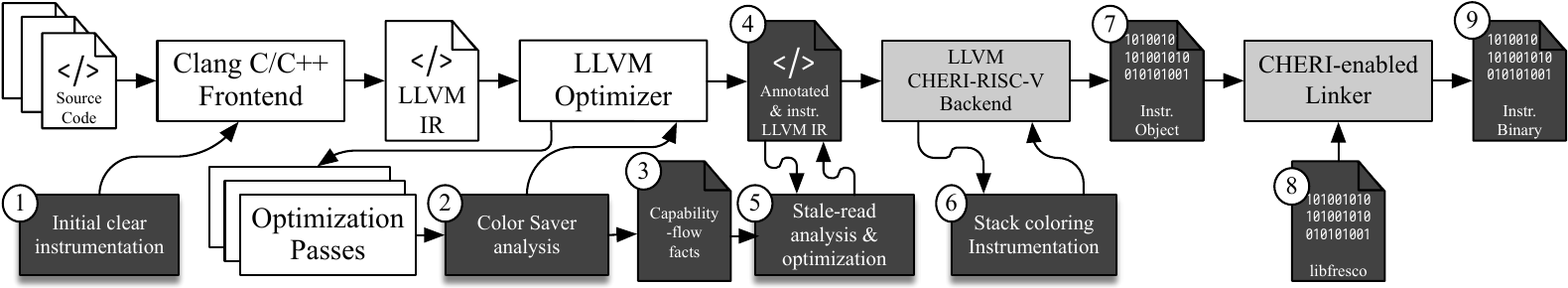}
    \caption{\FRESCO compiler architecture. Dark grey denotes additions or modifications needed to support stack coloring, while light grey indicates additions common to \gls{cheri}-enabled toolchains. White denotes components without significant changes.}\label{fig:compilerarch}
\end{centering}
\vspace{-0.2cm}
\end{figure*}

\subsubsection{Addressing Color Churn.}\label{para:churn}
In \PICASSO, a process's supply of colors is finite, bounded by the number of \gls{otype} bits reserved for the \ccpid.
For heap allocations this suffices in many workloads~\cite{Wang26}.
Stack frames, however, are allocated and retired orders of magnitude more frequently than heap objects are allocated and freed.
Applying \stackcoloring on the baseline \PICASSO design is impractical for any realistic workload.
We solve the color churn problem through two complementary mechanisms:
\begin{inparaenum}[1)]
    \item \emph{\ColorSaver} and
    \item \emph{\colorsegmentationlong}.
\end{inparaenum}

\paragraph{\ColorSaver.}\label{sec:color-saver}
Stack coloring prevents a capability derived from a frame from remaining
usable after that frame returns. \ColorSaver preserves this protection while
avoiding unnecessary coloring by answering two questions: 
\begin{inparaenum}[1)]
\item \emph{Can a capability pointing into the current
stack frame escape and remain usable after the function returns?}  This
happens when the function returns the capability, stores it in a global or heap
object, or passes it to a call whose target the analysis cannot resolve.
The escape analysis uses the answer to decide whether the function must be
colored.
\item \emph{Where can a capability left in reused stack memory be recovered?}
Stack allocation does not erase validity tags, so a capability loaded from a reused \emph{tag granule}---the capability-aligned, capability-sized unit with which \gls{cheri} associates a validity tag---can recover a \emph{stale capability} left there by an earlier use, provided no intervening store has cleared the tag.
\new{Because uncolored functions share their caller's color, such a capability may still bear a live color: its retraction waits on a return of the ancestor frame the color originates from, not on the frame that left the capability behind.}
The stale-read analysis therefore determines where the compiler
must clear tags before the program can recover stale capabilities.
\end{inparaenum}
\Cref{sec:color-saver-analysis} presents the algorithms and their guarantees, \Cref{sec:color-saver-implementation} their compiler implementation; the rules
appear in Appendix~\ref{app:escape-details} and~\ref{app:freshness-details} and the Rocq proof in the artifact.

\paragraph{\Colorsegmentationlong.}\label{sec:colorsegmentation}
\ColorSaver reduces how many functions draw colors, but cannot bound how many a thread consumes over time: long-running threads still exhaust any fixed pool, so the color churn problem persists.
\Colorsegmentationlong addresses color churn and \ref{goal:nocolorceiling}: rather than drawing from a single process-wide pool, it divides a processes' virtual memory into distinct \emph{color segments}, each with its own virtual \gls{pvt} and hence its own \ccpidhyph namespace.
\ifnotabridged
The scheme is analogous to memory segmentation, the \gls{os} technique that divides memory into distinct segments and identifies a location by a segment value and an offset within it.
\Colorsegmentation reverses this  relationship: the\else The \fi upper portion of a \cc's address \ifnotabridged (\dCFive in \Cref{fig:chericap})\fi identifies the segment the \ccpid belongs to. 
Because colors are scoped to their segment's address range, colors in one segment can never alias those in another. 

\paragraph{Scaling \stackcoloring with \colorsegmentation.}
\Stackcoloring dynamically divides the stack into disjoint address-space segments, each associated with its own \emph{color namespace}.
Frame colors are assigned from a segment's namespace until either its colors or memory is exhausted;
each exhaustion ends an \emph{epoch}, dividing the stack into successive epochs during which frames are backed by a particular segment.
\Stackcoloring tracks colors assigned within an epoch using a per-thread counter maintained by the instrumented prologues, and an 
\emph{epoch stub} in a runtime support library (\libfresco) manages segment transitions.
When the counter approaches the reserved range, the prologue invokes the stub (\dOne in \Cref{fig:stackcoloring}); the stub initializes a new stack segment (\dFour in \Cref{fig:stackcoloring}), switches the thread to it, and resets the counter.
Each thread can thus sustain \ccs across as many epoch switches as necessary, meeting ~\ref{goal:nocolorceiling}.
The number of segments a system supports is bounded only by available memory.
As the stack retracts, segments fall out of use; unused segments can be reclaimed and reused once a revocation sweep has invalidated capabilities pointing to them (\dFive in \Cref{fig:stackcoloring}).

\subsubsection{Compiler pipeline}

\Cref{fig:compilerarch} shows the compiler architecture supporting stack-coloring instrumentation. The Clang frontend lowers C/C++ to LLVM \gls{ir}, inserting an \initialclear at every variable that may hold a capability \dCOne. After the standard optimization pipeline, \ColorSaver classifies each function's stack allocations~\dCTwo, produces capability-flow facts~\dCThree, and annotates the \gls{ir}~\dCFour, identifying the functions from which a capability to the stack may escape.
Instrumentation then proceeds in two steps: the stale-read analysis removes \initialclears that capability-flow facts prove unnecessary~\dCFive; and the backend's \stackcoloring instrumentation \dCSix emits the coloring prologue and epilogue for the functions \ColorSaver selected. 
The CHERI-RISC-V backend lowers the instrumented code, and emits instrumented objects \dCSeven, which are linked against \texttt{libfresco} \dEight, to produce the instrumented binary \dCNine.

\ifnotarxiv
\begin{figure}[t]
\begin{centering}
    \includegraphics[width=\columnwidth]{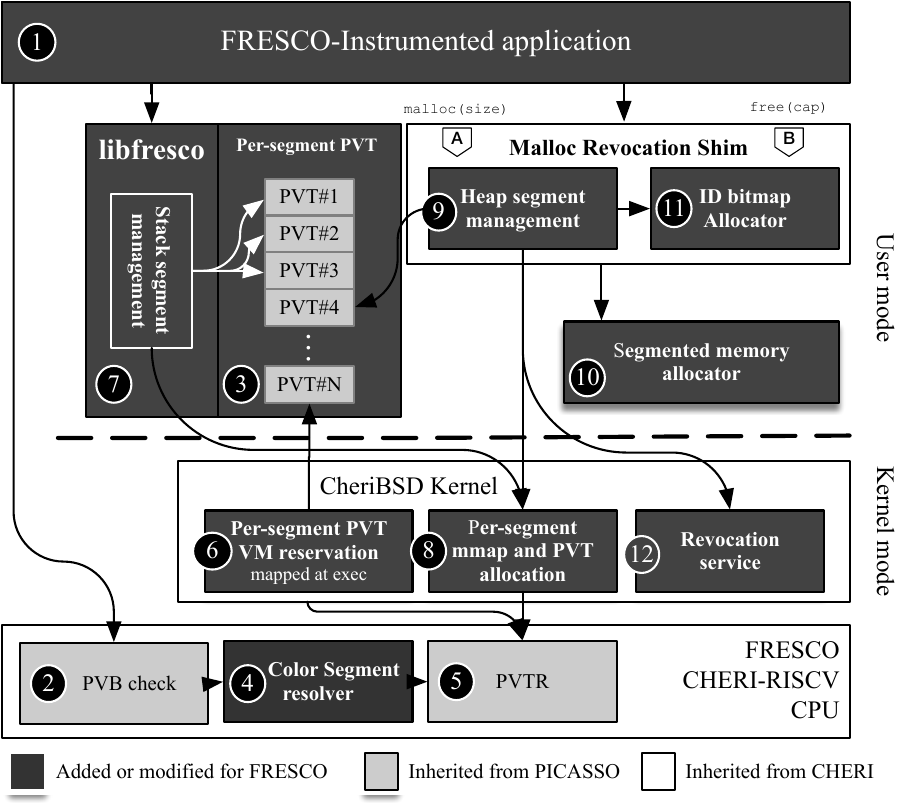}
    \caption{Overview of the \FRESCO runtime architecture. \ifnotabridged Dark grey denotes additions or modifications needed to support \stackcoloring and \colorsegmentation, while light grey indicates additions common to \PICASSO.\fi}\label{fig:runtimearch}
\end{centering}
\vspace{-0.2cm}
\end{figure}
\fi

\subsection{Color-segmented allocator}\label{sec:segmentedalloc-design}

\Colorsegmentation also benefits heap temporal-safety enforcement: a segment-aware \gls{mrs} and heap allocator can scale beyond 2M simultaneously live allocations and further delay capability revocation sweeps.
Allocation and deallocation follow \PICASSO's: \texttt{malloc()} claims an identifier through the \gls{mrs} and colors the returned capability with \gls{ccsettype}; \texttt{free()} clears the corresponding \gls{pvb}, detecting double-frees in the process, and releases memory for reuse.

A color-segmented allocator associates heap regions with distinct color segments, each with its own \gls{pvt} and \ccpid namespace. This composes with allocators that already partition their address space, since a color-segment boundary can coincide with a partition the allocator already maintains; an allocation's segment follows from the address the allocator returns, at no additional bookkeeping cost. As on the stack, colors are drawn from a segment's namespace until either its colors or its memory is exhausted. The allocator uses both conditions to decide where to place new allocations.

Three differences distinguish this design from \PICASSO's.
First, identifiers come from per-segment namespaces rather than a process-wide pool, so \ifnotabridged the number of simultaneously distinguishable heap allocations is no longer capped by the width of the \gls{otype} field, and \fi identifier exhaustion becomes a per-segment rather than a process-wide condition (\ref{goal:nocolorceiling}).
Spreading allocations across segments rather than one global namespace also makes revocation sweeps correspondingly rarer (\ref{goal:nocolorceiling}).
Second, identifier bookkeeping uses a flat bitmap allocator (Appendix~\ref{appx:bitmapalloc}) in place of \PICASSO's unr allocator, eliminating the fragmentation-dependent growth that dominates PICASSO's worst-case footprint (\ref{goal:boundedmetadata}).
\ifnotabridged
Gülmez et al.~\cite{Gulmez26} attribute \PICASSO's omnetpp overhead (\Cref{sec:temporal-safety-gaps}) to this maintenance and space cost of \ccpid allocations by the \gls{unr} allocator once the identifier structure fragments.
\fi
Third, heap and stack enforcement coexist on one system: each draws colors from disjoint segments, so neither consumes the other's namespace (\ref{goal:full}).

\ifnotabridged
In \PICASSO, revocation sweeps are triggered by \ccpid exhaustion, rather than memory pressure, and widening the \ccpid is not practical: additional identifier bits must be taken from the capability representation, where they compete with other \gls{cheri} extensions, and inflate both the \gls{pvt} and the identifier bookkeeping (\Cref{sec:temporal-safety-gaps}) regardless of if those additional \ccpids are needed by a particular workload.
Color segmentation removes this coupling.
In \FRESCO, because a colored capability's segment is determined by the upper portion of its address, and colors in distinct segments can never alias, a process's usable identifier space grows with the number of color segments rather than with the width of the \gls{otype} field.
\fi

\ifarxiv
\begin{figure}[t]
\begin{centering}
    \includegraphics[width=\columnwidth]{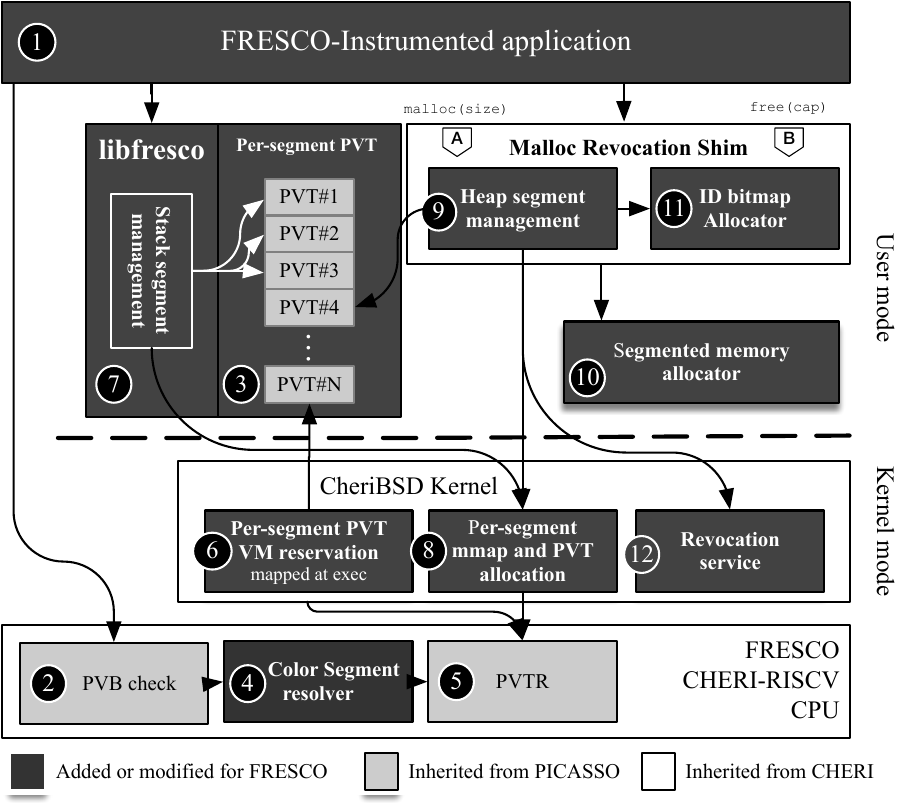}
    \caption{Overview of the \FRESCO runtime architecture. \ifnotabridged Dark grey denotes additions or modifications needed to support \stackcoloring and \colorsegmentation, while light grey indicates additions common to \PICASSO.\fi}\label{fig:runtimearch}
\end{centering}
\vspace{-0.2cm}
\end{figure}
\fi

\subsection{\FRESCO's runtime architecture}\label{sec:runtime-arch}

\Cref{fig:runtimearch} shows \FRESCO's architecture.
Every access through a \cc in a \FRESCO-instrumented application \dOne is subject to a hardware \gls{pvb}-check \dTwo, which confirms the capability's provenance \ifnotabridged is valid\fi before the access proceeds.
The check uses a per-segment \gls{pvt} \dThree. A \emph{color segment resolver} \dFour translates the \gls{pvt}-area base address held in the \gls{pvtr} \dFive, inherited from \PICASSO, into the correct per-segment \gls{pvt} offset, using the address of the capability\ifnotabridged responsible for the access\fi.
The \gls{pvt} area is reserved from process \gls{vmem} on process start\ifnotabridged (\texttt{execve})\fi, but physical pages are only mapped to a segment and its \gls{pvt} once the \emph{stack segment management} inside \libfresco\,\dSeven allocates the segment using a kernel-level facility \dEight.
For the heap, the \gls{mrs}' \emph{heap segment management} \dNine initializes heap color segments and directs the segmented memory allocator \dTen to use them, while the \gls{id} bitmap allocator \dEleven manages per-segment \ccpid allocations.
The in-kernel \emph{revocation service} \dTwelve, also inherited from \PICASSO, fully revokes stale capabilities for reclaimable segments reported by the stack \dFive and heap \dTen segment management. \Cref{sec:segmentedalloc} details the adjusted revocation process.

\section{\ColorSaver Analysis}\label{sec:color-saver-analysis}

\subsection{Escape Analysis}\label{subsec:escape-analysis}

The escape analysis traces possible capability flows backward
from where a capability could outlive the function:
return values, globals, and arguments passed to calls that cannot be resolved.
To do so, it summarizes capability-relevant LLVM instructions and repeatedly
propagates possible flows through copies, loads, stores, and calls.  A function must be colored if a
capability pointing to its frame may escape; otherwise, its coloring prologue and epilogue can be omitted.

\paragraph{Capability-flow facts.}
The analysis represents explicit capability flow with three kinds of facts (adopted from~\cite{Esswood21}):
\vspace{-6pt}
\[
  \mathsf{Copy}(d,s),\qquad
  \mathsf{Load}(d,p),\qquad
  \mathsf{Store}(p,v),
  \vspace{-6pt}
\]
where $d$, $s$, $p$, $v$ are LLVM values.
They state, respectively, that a capability in $s$ may flow to $d$, that
$d$ may receive a capability read through $p$, and that a capability in $v$
may be written through $p$.  
Casts, address calculations, \texttt{phi} nodes,
and \texttt{select} instructions preserve capabilities and  produce copy facts. Capability loads and stores produce load and store facts.

\paragraph{Call graph.}
For direct calls, the analysis connects its
actual arguments to formal parameters and its returned values to the call
result, allowing capability flows to cross function boundaries. Otherwise, it uses the call's summary where
available, or falls back to treating the call's arguments and result as possible escape
routes. Globals, the current function's arguments and returns, inline
assembly, and unsupported capability operations are handled similarly.

\paragraph{Propagation.}
Tracking only escaping LLVM values is insufficient, since a capability
may be stored in memory and later recovered through another pointer.  The analysis follows both direct and memory flow, using a finite,
conservative abstraction that may color a function
unnecessarily, but cannot hide a possible escape.
Starting from possible escape routes, the solver repeatedly applies the propagation rules until no new flow is found.  The finite
LLVM input and memory-flow abstraction ensure termination.  A function is then colored if a capability pointing into its frame 
reaches a possible escape route.  Detailed propagation rules are given in
Appendix~\ref{app:escape-details}.

\paragraph{Guarantee.}
Assume that extraction records every way a capability can be copied, loaded,
stored, passed across a call, or exposed outside the analyzed modules, and that no capability pointing into a new frame exists before that frame
is created.  Under these assumptions, the solver has no false negatives: if it
omits coloring for a function, no capability pointing into that function's
frame can escape.
We mechanized this result for the propagation algorithm in Rocq.  The proof
does not verify the LLVM extraction code nor the optimizations in Appendix~\ref{app:escape-optimization}; the end-to-end guarantee depends on extraction covering every relevant LLVM construct.
Appendix~\ref{app:escape-proof} describes the guarantee; the artifact has the mechanized proof.

\subsection{Stale-Read Analysis}\label{subsec:uninitialized-read}

Stale-read analysis prevents a stale capability left by an earlier use of stack
storage from being recovered after that storage is reused: it zeroes the
storage of every declared local variable that could hold a capability, then
removes those clears that the program's own stores already render redundant.

\paragraph{Variables that may hold capabilities.}
Recall that \gls{cheri} holds one validity tag per naturally-aligned, capability-sized granule (16 bytes for 128-bit capabilities), out of band from the data itself (\Cref{sec:cheri}). The tag is set only by a capability-width store of a tagged value, and cleared by any overlapping non-capability store.
Tag-preserving bulk writes, e.g. \texttt{memcpy()}) whose destination is declared as bytes or capability-carrying integer types in the pure-capability \gls{abi}, can propagate capabilities into opaque storage under \Adv influence.
Even storage that never receives a capability during its own lifetime may sit on granules that a previous occupant of that frame slot left tagged.
Recovery under our adversary model (\Cref{sec:system-and-adversary-model}) only requires that the granule survive overwriting until an \Adv-influenced path issues a capability load from it.
To recover a stale capability, \Adv must cause stack storage to be reused and
read as a capability before any store clears or replaces its tag. Since
\Adv can neither forge a capability nor widen its bounds, a compiler-generated
slot is accessible only if the compiler exposes a capability whose bounds
include it.  Under the trusted-compiler assumption
(\ref{asm:comp-is-trusted}), every such slot is written before any capability
load from it.  Storage for a declared local variable, however, can be read
before initialization along a path selected by \Adv, triggering undefined behavior exposing stale capabilities.
We therefore identify  declared local variables that can propagate stale capabilities.

Stale-read analysis selects every local variable with automatic storage duration,
excluding parameters and variables of C++ reference type, whose type is
capability-carrying or recursively contains one through an atomic value,
array element, structure or union member, or C++ base or member. It also
selects variables at least one tag granule in size, all variable-length arrays,
and integers explicitly cast to a capability type.  The size-based cases cover
type punning and copies that preserve validity tags. Stale-read instrumentation
aligns and zeroes the storage of selected variables at their definition,
before any initialization or use; this zeroing is the \emph{\initialclear}.

\paragraph{Removing unnecessary initial clears.}
A \emph{possible read} of a tag granule is a capability load overlapping the
granule, or a tag-preserving memory copy whose source overlaps it.
A \emph{definite overwrite} is a store other than the initial
clear that overlaps the granule; a non-capability store clears the old tag and a capability store
replaces it.  If the analysis cannot resolve
an operation's target, offset, size, or effect, it records a
possible read for every tag granule of the variable and no definite overwrite.
An \initialclear is removed if every path from the variable's
definition to a possible read contains a definite overwrite before the read;
otherwise, the clear is retained.
Appendix~\ref{app:freshness-details} describes the details.

\paragraph{Guarantee.}
Assume that stale-read analysis
\begin{inparaenum}[1)]
\item selects every declared local from whose storage a stale capability can be
recovered; 
\item every operation that can recover or propagate one is classified
as a possible read;
\item the control-flow model includes every executable path; and
\item a definite overwrite is recorded only for a store guaranteed to clear
or replace the tag.
\end{inparaenum}
Under these assumptions, the analysis is sound: every stale capability read is preceded by either a retained initial
clear or definite overwrite, so \Adv cannot recover stale capabilities
by the program reading reused storage before
writing it.

\section{FRESCO Implementation}
\label{sec:implementation}

We implement \FRESCO for CHERI-RISC-V in the QEMU full-system emulator and CHERI-Toooba \gls{fpga} (\Cref{sec:hardware}). We also extend the \gls{cheri}-enabled Clang/LLVM compiler with \stackcoloring and \ColorSaver (\Cref{sec:compiler}), and add the \libfresco runtime, segment-aware \gls{mrs}, allocator, and revocation to CheriBSD (\Cref{sec:cheribsd}).

\subsection{\FRESCO CHERI RISC-V Hardware}\label{sec:hardware}
\paragraph{Per-segment \glspl{pvt}.}
\FRESCO extends \PICASSO with address-based \colorsegmentation, allowing the same color value in different memory segments without aliasing in the \gls{pvt}. Recall from \Cref{sec:colorsegmentation} that each segment has its own virtual \gls{pvt} placed in a contiguous \gls{pvt} reservation area beginning at a base address $B$ configured to the \gls{pvtr} \gls{csr}.
\FRESCO describes the location of a segment's \gls{pvt} to the hardware through three \glspl{csr}: the table base $B$, the segment shift $s$ (so a segment spans $2^{s}$ bytes), and the segment count $N$ (a power of two).
A capability with address $a$ and \gls{otype} $t$ then addresses bit $t \bmod 8$ of the \gls{pvt} byte
\vspace{-6pt}
\[
  B \;+\; \underbrace{\big((a \gg s)\mathbin{\&}(N-1)\big)}_{\mathrm{segment}}\, Z \;+\; \lfloor t/8 \rfloor,
  \vspace{-6pt}
\]
where $Z$ is the per-segment slice size.
The three \glspl{csr}---$B$, $s$, and $N$---are programmed per process at startup from parameters embedded into the \gls{elf}, or from built-in defaults, so the segment size and count are configurable rather than fixed in hardware; only $Z$, tied to the \gls{otype} width, is constant.
The address selects the segment and the \gls{otype} the bit, so the same \gls{otype} in two segments resolves to different \gls{pvt} bytes---a per-segment namespace with no software involvement. Heap and stack share this \gls{pvt} addressing scheme, so neither needs to know which kind of segment a capability belongs to.

\paragraph{Retracting capabilities.}
Retracting \ccs means invalidating their \gls{pvb}.
In \PICASSO this is a read-modify-write of a single bit that races with any other core doing the same to a neighboring bit in the same word.
In software it costs a load, an \texttt{or}, a store, and a fence on every deallocation (\Cref{sec:cheri-temporal-safety}).
\FRESCO adds a \texttt{csealtype} instruction that sets a \gls{pvt} bit atomically in a segment's \gls{pvt}, decoded and issued through the memory pipeline like other atomic operations, which removes both the fence and the window in which a concurrent retire could be lost.
Additionally, \texttt{csealtype} selectively invalidates the cached \gls{pvb} of the \gls{pvt}-buffer, avoiding the need to flush the entire buffer using a fence.
\texttt{csealtype} takes the capability whose color is being retired as operand and uses the address-derived indexing described above;  software need not compute a table offset.

\subsection{\FRESCO Clang/LLVM Compiler}\label{sec:compiler}

\subsubsection{\ColorSaver Analysis}\label{sec:color-saver-implementation}

\ColorSaver runs at a late LLVM \gls{ir} boundary: after the 
optimization pipeline but before instruction selection. This \gls{ir} still
carries capability types, call attributes, and stack-object lifetime markers,
yet already reflects the optimizer's transformations. The escape
classifier marks the functions the backend must emit stack-coloring
instrumentation for; the stale-read analysis rewrites the \gls{ir} with validity tag clears.

\paragraph{Whole-program escape classification.}
Separate compilation normally hides definitions in other translation
units from an LLVM module pass, forcing calls to them to be
treated as opaque. We therefore add an analysis compilation that reduces every
optimized module in the analyzed build to the capability-flow facts of \Cref{subsec:escape-analysis} and solves over the combined corpus. The solver matches compatible declarations and definitions across
module boundaries and computes call summaries and escape classifications in
a single global fixed point, so a caller can use a precise callee summary
regardless of which translation unit the callee came from. Missing,
ambiguous, and interposable targets remain conservatively modeled as external
calls. We then recompile with the resulting per-function classifications
attached to the corresponding LLVM functions, before backend instrumentation.

\paragraph{Stale-read clearing.}
We modified the Clang frontend to emit an \texttt{llvm.cheri.capability.auto.init} intrinsic at each
eligible stack object definition. The stale-read analysis, implemented alongside the escape analysis, then removes unnecessary clears. Each remaining intrinsic
is lowered to an LLVM memory-set operation that zeroes its
range. Instruction selection expands this operation into a \gls{cheri} target's zeroing
sequence which clears capability tags.

\begin{figure}[t]
    \centering
    \begin{tikzpicture}[
        font=\small,
        >=Latex,
        frame/.style={
            draw=black!70,
            minimum width=3.0cm,
            minimum height=0.50cm,
            align=center,
            inner sep=2pt
        },
        phase/.style={
            frame,
            fill=blue!12,
            draw=blue!60!black
        },
        action/.style={
            frame,
            fill=orange!18,
            draw=orange!70!black
        },
        body/.style={
            frame,
            minimum height=1.00cm,
            fill=gray!10
        },
        explanation/.style={
            draw=black!55,
            text width=3.5cm,
            align=left,
            inner sep=5pt,
            fill=white
        },
        callout/.style={
            ->,
            thick,
            draw=black!65
        }
    ]
    \node[frame, fill=gray!28] (caller) {
        Function entry
    };

    \node[phase, below=0pt of caller] (phaseA) {
        \textit{Phase A}
    };

    \node[action, below=0pt of phaseA] (allocate) {
        Frame Allocation
    };

    \node[phase, below=0pt of allocate] (phaseB) {
        \textit{Phase B}
    };

    \node[body, below=0pt of phaseB] (function) {
        Function body
        
    };

    \node[action, below=0pt of function] (epilogue) {
        Epilogue
    };
    \begin{scope}[on background layer]
        \node[
            draw=white!70,
            thick,
            fit=(caller) (epilogue),
            inner sep=6pt, label={[font=\bfseries]above:Instrumented colored call}
        ] {};
    \end{scope}
\draw[->, very thick, draw=black!60]
    ([xshift=-0.40cm]caller.west)
    -- ([xshift=-0.40cm]epilogue.south west)
    node[midway, left=8pt, rotate=270, anchor=center]
    {\scriptsize execution order};
\node[explanation, right=0.8cm of phaseA] (explainA) {
    \dOne Read the caller's color $\ell$\\
    \dTwo Increment $k$ to $k+1$\\
    \dThree Check whether $k<K$\\
    \dFour Recolor \gls{csp} with $k+1$
};
    \node[explanation, right=0.8cm of phaseB] (explainB) {
    \dFive Save the caller's color $\ell$
    };

    \node[explanation, right=0.8cm of epilogue] (explainE) {
        \dSix Restore $\ell$\\
        \dSeven Retract color $k+1$
    };
    \draw[callout] (phaseA.east) -- (explainA.west);
    \draw[callout] (phaseB.east) -- (explainB.west);
    \draw[callout] (epilogue.east) -- (explainE.west);

    \end{tikzpicture}

    \caption{
        Prologue and epilogue of an instrumented colored function.
        Phase~A executes before frame allocation and performs the operations
        that must precede a possible epoch switch. Phase~B records the
        caller's color after the frame and callee-save area have been
        established. $k=$ segment's epoch counter; $K=$ threshold; $\ell=$ caller's \ccpid.
    }
    \label{fig:colored-prologue}
\end{figure}
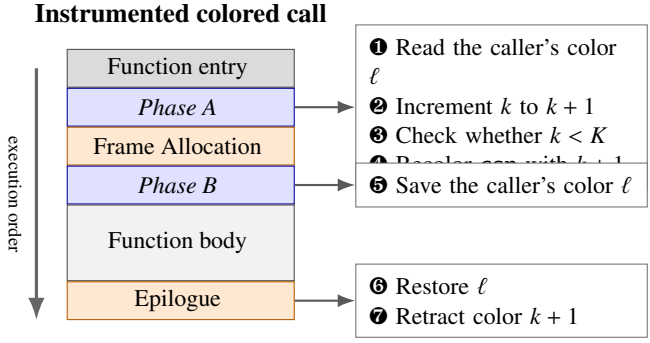

\subsubsection{\StackColoring Instrumentation}
\label{sec:instrumentation}

The \texttt{-fcheristackcolor} flag enables \stackcoloring instrumentation; \FRESCO applies it only to functions the escape analysis (\Cref{subsec:escape-analysis}) selects.
Each colored frame receives a fresh color from a monotonically
increasing counter $k$  held in \gls{tls}.
\FRESCO splits the prologue around frame allocation into \emph{Phase~A} and \emph{Phase~B} (\Cref{fig:colored-prologue}).
Phase~A precedes frame allocation: it reads the caller's color out of \gls{csp} \dOne, increments the color counter to $k+1$ \dTwo, and compares the result against the epoch threshold ($K$) \dThree.
If the threshold has not been reached, \emph{Phase~A} recolors \gls{csp} \dFour and continues on the common path; otherwise, it invokes the \texttt{libfresco} epoch stub (\Cref{sec:libfresco}) which switches the thread stack to a new segment and retries coloring.
This ordering ensures the frame is not written to before a segment switch.
Phase~B follows frame allocation and callee-save spills, and performs stores that require a frame.
It writes the caller's color in a reserved slot \dFive, which remains until the epilogue restores it \dSix.

\paragraph{\Gls{abi} compatibility.}\label{sec:varargs}
A segment switch moves \gls{csp} to a different address range, which affects both stack-passed arguments and the address used by \texttt{va\_start} to support functions with variadic arguments.
The prologue and epoch stub preserve the \gls{abi} transparently to the
instrumented function.
The RISC-V \gls{abi} (\Cref{sec:cheri}) passes eight integer arguments in \texttt{a0}--\texttt{a7} and places additional arguments in the caller's outgoing-argument area.
A callee accesses these arguments at positive offsets from its entry \gls{csp}.
Functions with more than eight arguments, and variadic functions, therefore read part of their parameter list through
\gls{csp}, and after a segment switch those offsets would refer to the new segment rather than to the original argument area.
Before invoking the epoch stub, Phase~A records the size of the argument area into \gls{tls}; when the stub switches segments, it reserves an equivalent area below the new segment's top and copies the arguments across.
The destination derives from the new segment's capability, so \gls{cheri}'s bound checks prevent the copy from exceeding the segment.
The work is proportional to the argument-area size and is paid only on the overflow path.
For variadic functions, Phase~A additionally saves the incoming \gls{csp} in the
thread-local context; because the stub clobbers caller-saved registers,
Phase~B reloads this value once the frame exists and makes it available to
\texttt{va\_start}.
Its location is fixed by an agreement between the backend and \libfresco.

\subsection{\FRESCO CheriBSD}\label{sec:cheribsd}

\subsubsection{Kernel segment management}\label{sec:kernel}

At \texttt{execve()} the kernel reserves a contiguous virtual memory region, the \emph{segment pool}, together with the corresponding PVT reservation area, sizing both from the \gls{elf}-embedded segment count and segment size, or from built-in defaults when those parameters are absent.
The same parameters program the three segment-addressing \glspl{csr} of \Cref{sec:hardware}, so hardware and runtime agree on a process's segment layout.
Both regions are passed to userspace as bounded capabilities.
Physical pages back a segment and its \gls{pvt} slice only when that segment is first mapped through the kernel's segment-allocation facility, so unused segments cost virtual address space rather than memory.
The in-kernel revocation service sweeps asynchronously, fully revoking capabilities into the segments that the stack and heap segment managers report as reclaimable.

\subsubsection{Stack segment management in \texttt{libfresco} runtime}\label{sec:libfresco}\label{sec:epochstub}

\begin{figure}[t]
    \centering
    \begin{tikzpicture}[
        font=\footnotesize,
        seg/.style={draw, rounded corners, minimum width=1.3cm,
                    minimum height=1.1cm, align=center, inner sep=1pt},
        fwd/.style={-{Latex[length=1.5mm]}, thick},
        bwd/.style={-{Latex[length=1.5mm]}, thick, dashed},
        lbl/.style={font=\scriptsize\itshape, inner sep=1pt}
    ]
        \node[seg, fill=blue!15]  (s1) at (0,0)   {Seg\,1\\ {\scriptsize\texttt{depth}\,0}};
        \node[seg, fill=red!15]   (s2) at (1.9,0) {Seg\,2\\ {\scriptsize\texttt{depth}\,1}};
        \node[seg, fill=green!15] (s3) at (3.8,0) {Seg\,3\\ {\scriptsize\texttt{depth}\,2}};
        \node[seg, fill=orange!20](s4) at (5.7,0) {Seg\,4\\ {\scriptsize\texttt{depth}\,3}};
        \node (dots) at (7.1,0) {$\cdots$};

        \draw[fwd] (s1.north) to[out=55,in=125] (s2.north);
        \draw[fwd] (s2.north) to[out=55,in=125] (s3.north);
        \draw[fwd] (s3.north) to[out=55,in=125] (s4.north);
        \draw[fwd] (s4.north) to[out=55,in=125] (dots.north);

        \draw[bwd] (s2.south) to[out=-125,in=-55] (s1.south);
        \draw[bwd] (s3.south) to[out=-125,in=-55] (s2.south);
        \draw[bwd] (s4.south) to[out=-125,in=-55] (s3.south);

        \node[lbl] at (3.3,1.25) {\textbf{\dOne Stack Segment Switch}: \texttt{depth}\,+1, color new \gls{csp}};
        \node[lbl] at (2.85,-1.55) {\textbf{\dTwo Return trampoline}: \texttt{depth}\,$-1$, restore prior segment};
    \end{tikzpicture}
    \caption{Stack segment nesting. Each time $k$ reaches its pre-set threshold, \libfresco moves the thread onto a fresh segment pushed onto the segment stack; returns emptying a segment pop it. Segments are allocated lazily with an own $k$.}
    \label{fig:epochstack}
\end{figure}
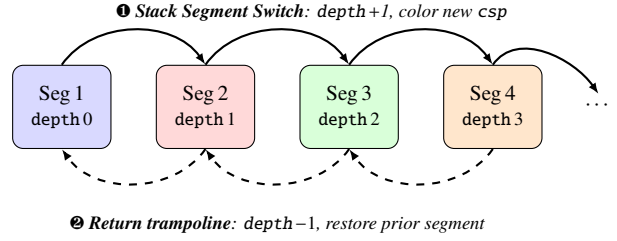

\Stackcoloring eventually exhausts the $2^{21}$ color values encoded in the \gls{otype} (\Cref{sec:stackcoloring}). \FRESCO therefore partitions the stack into disjoint color segments (\Cref{fig:epochstack}), each with an independent
color namespace. When a segment's color counter reaches its threshold $K$ (Phase A in \Cref{fig:colored-prologue}), the epoch
stub supplied by \texttt{libfresco}, the runtime support library introduced in \Cref{sec:runtime-arch}, switches the stack to a fresh segment by saving the current \gls{csp}, return address, and counter; and  installing a return trampoline as the triggering function's return address.
When the function which triggered a segment switch returns, the trampoline saves the retiring segment's color counter to its \gls{tls} entry, restores the caller's \gls{csp}, return address, and color counter, decrements \texttt{depth}, and returns.

\paragraph{Epoch stack.}
\FRESCO records the current segment nesting depth and, for each active segment,
stores the previous \gls{csp}, return address, and color counter in a
\gls{tls} descriptor. These descriptors form a dynamically grown
doubly-linked list. On return, the \libfresco runtime uses the descriptor to restore
the previous segment. If a segment exhausts its color space, \texttt{libfresco}
allocates a new segment for that nesting level and resets its counter.
Thus, neither nesting depth nor the color space of an individual segment
imposes a fixed bound.

\paragraph{Multi-threading.}
\libfresco interposes on thread creation: it allocates a segment from the
\FRESCO segment pool and installs it as the new thread's stack, so that every
thread colors frames against its own \gls{pvt}.
The interposer also wraps the start routine so that the thread begins with the
epoch counter $k = 0$, the same baseline the main thread receives at process
start.  \libfresco registers a thread destructor the first time a thread
allocates a segment; at thread exit it walks the segment list and returns each
segment to the shared pool (\Cref{sec:segmentedalloc}). \libfresco applies the same principle at \texttt{fork}: the child inherits the parent's counter and \gls{pvt} copy-on-write, so a child handler leaves $k = K$
and lets the ordinary overflow path give it a fresh segment with a clean
\gls{pvt}.

\paragraph{Non-local control flow.}
Stack unwinding and \texttt{longjmp} resume execution in an outer frame without
running the epilogue of the function they leave. \libfresco{} therefore exposes a recovery routine that takes the stack
address of the target frame, retracts the colors of every frame deeper than
the target, and restores the latest $k$. C++ exception handling
invokes the recovery routine from the personality routine before each frame runs its cleanup
landing pad since destructors executed during an unwind are themselves
colored calls, and again on entry to the catch handler. \texttt{longjmp} needs a single invocation, using the saved \gls{csp} in the \texttt{setjmp} jump buffer.

\subsubsection{Heap segment management in \gls{mrs} and allocator}\label{sec:segmentedalloc}

The \gls{mrs} activates segments from a pool reserved at initialization, so activation requires no system call or memory allocation. Colors and memory are tracked separately. On \texttt{malloc()}, the \gls{mrs} checks whether the current segment's colors are nearing exhaustion, and if so, activates a new segment or queues revocation. When the underlying allocator cannot serve a request, the \gls{mrs} attempts to switch to a segment with memory available. In either case \gls{mrs} prefers a reusable segment---one whose colors revocation has reclaimed, or one in which the underlying allocator has since freed memory---and activates a fresh one only if none is found.

The allocator then reserves memory from the selected segment, or non-segmented memory via \texttt{mmap()} for allocations exceeding the segment size. \FRESCO identifies the segment containing the returned address and assigns a color from that segment's bitmap; addresses in non-segmented memory draw from a global bitmap and \gls{pvt}.  Every allocation thus receives a unique color regardless of size, and the segments need not accommodate arbitrarily large objects. That fallback namespace is finite, but with the default 4\,MB segment size an application would need more than $\approx~8$\,TiB of live oversized allocations to exhaust it, and larger segments can be configured.

On \texttt{free()}, \FRESCO identifies the segment and sets the corresponding \gls{pvb}---or the \gls{pvb} in the global \gls{pvt} for non-segmented addresses---retracting the capability. The underlying allocator then releases the memory for immediate reuse.
\Cref{fig:segmentedmalloc,fig:segmentedfree} in Appendix~\ref{appx:mallocfree-figs} illustrate both flows.

\paragraph{Triggering revocation.}
\FRESCO triggers revocation proactively when the free segment pool falls below a threshold, or when no segment can be reused or freshly activated. The sweep runs asynchronously (\Cref{sec:kernel}), and each \texttt{malloc()} polls for its completion so colors are reclaimed as soon as the kernel sweep finishes. As in \PICASSO, reclamation snapshots the \gls{pvt}, clears the corresponding bits from the live \gls{pvt}, and returns the snapshotted colors back to the \gls{id} allocator; in \FRESCO, this procedure is applied per segment.

\paragraph{Color density and segment utilization.}
A segment retires when it exhausts either its memory or its colors, and the two are consumed by different quantities: memory tracks the live heap footprint; colors, the cumulative allocation count, since freed memory returns to the allocator immediately but its color only at the next sweep. Workloads of small, short-lived allocations therefore churn colors independently of their working-set size and can exhaust a segment's colors while much of its memory remains recyclable. Such a segment retires early: its memory sits idle until a sweep reclaims the colors, its 512 KiB of metadata is amortized over less useful memory, and the pool drains faster, triggering revocation more often. Because  the per-segment colors and metadata are fixed, segment size governs this balance: larger segments amortize metadata over more memory but waste more of it when colors run out first; smaller ones track color-hungry workloads more closely at the cost of more segments and more aggregate metadata. \ifnotabridged\FRESCO defaults to 4 MiB segments and a $2^{21}$-color namespace.\fi

\paragraph{Jemalloc integration.}
Our prototype builds on jemalloc~\cite{Evans26} which does not partition its address space into fixed regions. \FRESCO binds each segment to a dedicated jemalloc arena and modifies the arena's extent allocation to draw from the segment's address range: it first attempts to retrieve memory from the segment's pre-allocated region via an atomic bump pointer, falling back to a \texttt{mmap()} path for allocations exceeding the segment size. Only this slow path changes. The fast path uses tcache and size-class pools, continuing to recycle memory within the segment unchanged, preserving jemalloc's cache behavior. A segment switch therefore need not flush the tcache: when the tcache serves recycled memory from a previous segment, the address-based lookup identifies the segment and assigns a color from its namespace. Only when those colors are fully exhausted is the  allocation retried on the current segment. Retries, however, can thrash the cache: in the worst case every cached entry belongs to an exhausted segment, and each must be returned to its arena before  a fresh allocation is requested, draining tcache one entry at a time.

\ifnotabridged
\paragraph{Memory overhead.} The per-segment overhead consists primarily of the otype bitmap and the \gls{pvt}, each 256\,KiB, totalling 512\,KiB per segment. Additionally, each segment is bound to a jemalloc arena, which brings its own metadata overhead (bin structures, extent trees, etc.). The exact size of this per-arena metadata depends on jemalloc's configuration but is small relative to the segment's 4\,MiB memory region. As a consequence, the segment count represents another trade-off. More segments delay revocation, but increase the memory overhead. Fewer segments reduce the memory overhead,  but require more frequent revocation cycles to reclaim exhausted colors.

\paragraph{Interaction with stack coloring.} Heap and stack coloring share a segment pool, allocated by the kernel at process startup. Both draw memory segments from the pool, but their usage differs. Heap segments are bound to jemalloc arenas, and otype allocation and assignment is handled by the \gls{mrs}. For stack segments, the \gls{mrs} exposes an \gls{api} that hands the caller bounded capabilities to the segment's memory and its corresponding PVT, leaving otype management entirely to the stack coloring runtime. A unified state tracker ensures that neither side claims segments owned by the other. Retired stack segments are reclaimed during the same revocation sweep that reclaims heap otypes. This design requires no static partitioning of the segment pool. Both consumers draw from a shared pool, and segments released by one can be reused by the other after a revocation cycle.
\fi

\section{Evaluation}\label{sec:evaluation}

\begin{table}[t!]
    \centering
    \caption{Area costs of \FRESCO and \PICASSO on the CHERI-Toooba for VCU118 @ 25 MHz compared to CHERI-Toooba.}\label{tab:hwcost}
    \resizebox{\columnwidth}{!}{
    \Large
    \begin{tabular}{rc rcc rcc}\toprule
    & \multicolumn{1}{c}{\textbf{CHERI-Toooba}}
    & \multicolumn{2}{c}{\textbf{\PICASSO}}
    & \multicolumn{2}{c}{\textbf{\FRESCO}} \\
    & \textbf{value} & \textbf{value} & \textbf{$\Delta$ (\%)} & \textbf{value} & \textbf{$\Delta$ (\%)} \\ \midrule
    LUTs      & {688096}  & {720347}  & {32251(+4.69)}  & {722488} & {34392(+5.00)} \\
    Memory    & {20113}   & {21611}   & {1498(+7.65)}   & {21611}  & {1498(+7.65)} \\
    Registers & {419300}  & {445129}  & {25829(+6.16)}  & {446164} & {26864(+6.40)} \\
    \bottomrule
    \end{tabular}}
\end{table}

We evaluate \FRESCO's security, performance, and hardware area cost using the BESSPIN-GFE evaluation platform. We replace the standard CHERI-Toooba core with our \FRESCO CHERI-Toooba, an extension of the CHERI-RISC-V Toooba \gls{fpga} softcore (RV64ACDFIMSUxCHERI) built on the open-source Bluespec RISC-V 64-bit Toooba core. 
We synthesize the \gls{soc} at 25MHz (BESSPIN-GFE default) for the Xilinx Virtex UltraScale+ VCU118 FPGA. We conduct the security evaluation (\Cref{sec:seceval}) both on the \gls{fpga} platform and a functionally equivalent \texttt{QEMUsystem-CHERI128} full-system emulator. We bound \FRESCO at 2048 4-MiB segments.  We answer the following research questions:
\begin{enumerate}[label=\textbf{RQ\arabic*},labelsep*=0pt]
\item:~{Can \FRESCO detect use-after-return and use-after-free?}\label{rq:security}
\item:~{How does \ColorSaver impact color churn?}\label{rq:colorsaver}
\item:~{What is \FRESCO's hardware cost and performance?}\label{rq:non-cc-performance}
\item:~{What is \FRESCO's run-time and memory overhead?}\label{rq:cc-heap}\label{rq:cc-performance}
\end{enumerate}

\subsection{Security Evaluation}\label{sec:seceval}

To answer \ref{rq:security}, we evaluate \FRESCO on the Juliet Test Suite 1.3, using the CWE-415 (double free), CWE-416 (use after free), and CWE-562 (return of stack variable address). Across 1,214 vulnerable (``bad'') and patched (``good'') test cases, \textbf{\FRESCO detects all bad cases with zero false positives}.

We further evaluate \FRESCO against real-world CVEs, GitHub issues, and OSS-Fuzz reports\ifnotabridged, which currently list 117 issues tagged \emph{stack-use-after-return}\fi. On the heap, \FRESCO prevents exploitation of BZip2 (CVE-2016-3189), nasm (CVE-2017-10686, CVE-2019-8343), libzip (CVE-2019-17582), NGINX (CVE-2020-24346), LibreDWG (CVE-2022-35164), Lua (CVE-2019-6706), mjs (issues 73 and 78), and yasm (issue 91), reproduced from the test suites in ~\cite{Nguyen20,Ahn24}. On the stack, \FRESCO detects use-after-return in nasm (CVE-2020-21686)~\cite{Suhwan20}, mjs (issue 301)~\cite{Vancir24}, and libucl (OSS-Fuzz \#42500636)~\cite{OSSFuzz21}. We also evaluate the \gls{poc} from Microsoft's analysis of the CHERI ISA~\cite{Joly20}, a complete exploit chain against JavaScriptCore (JSC) exploiting a dangling stack pointer, and confirm that \FRESCO detects the access to the retired stack frame. None of these use-after-returns are discovered by the compiler (\Cref{sec:relatedwork}), but \FRESCO detects them in all the reproduced cases.
Across all experiments, \textbf{\FRESCO demonstrates complete coverage of the evaluated temporal-safety violations across heap and stack}. These results support the case that \FRESCO addresses concerns of  memory-safety completeness (\Cref{sec:discussion}).

\subsection{Color Saver Analysis}\label{sec:colorsavereval}
\begin{table*}[t]
\centering
\caption{\ColorSaver (CS) vs. Full coloring (Full): Stack and Heap segment and revocation statistics (SPEC INT2006).}
\label{tab:decision-vs-full}
\resizebox{\textwidth}{!}{
\begin{tabular}{l r rr rr rr rr rr r c}
\toprule
& & \multicolumn{2}{c}{Stack segments} & \multicolumn{2}{c}{Fast path calls} & \multicolumn{2}{c}{Slow path calls} & \multicolumn{2}{c}{Max depth} & \multicolumn{2}{c}{Revocations} & & \\
\cmidrule(lr){3-4} \cmidrule(lr){5-6} \cmidrule(lr){7-8} \cmidrule(lr){9-10} \cmidrule(lr){11-12}
Benchmark & Safe / Unsafe & Full & CS & Full & CS & Full & CS & Full & CS & Full & CS & \# Allocations & Heap segments \\
\midrule
xalancbmk   & 15287 / 1245 & 6460 & 460 & 1.35e10 & 9.52e8 & 8.38e6 & 2.83e6 & 8  & 6  & 4 & 0 & 1.06e6 & 114 \\
omnetpp     & 2085 / 111   & 3470 & 2   & 6.86e9  & 3.01e6 & 4.03e8 & 9.84e5 & 3  & 1  & 2 & 0 & 1.30e8 &  63 \\
omnetpp (ref) & 2085 / 111 & 7740 & 10  & 1.54e10 & 9.77e6 & 8.31e8 & 7.29e6 & 3  & 2  & 4 & 3 &  2.67e8 &  95 \\
h264ref     & 504 / 13     & 2820 & 1   & 5.60e9  & 4      & 3.00e8 & 0      & 4  & 0  & 1 & 0 & 3.83e4 & 10 \\
sjeng       & 111 / 19     & 2170 & 88  & 4.53e9  & 1.77e8 & 7.66e5 & 7.22e3 & 10 & 6  & 1 & 0 & 5      & 2 \\
bzip2       & 59 / 4       & 239    & 3   & 3.74e8  & 18     & 1.16e8 & 0      & 3  & 0  & 0 & 0 & 87     & 4 \\
gobmk       & 2402 / 71    & 1530 & 8   & 3.16e9  & 1.68e5 & 5.03e5 & 0      & 8  & 0  & 0 & 0 & 1.21e5 & 9 \\
hmmer       & 455 / 21     & 40     & 1   & 8.09e7  & 6      & 5.17e5 & 0      & 2  & 0  & 0 & 0 & 1.70e5 & 3 \\
libquantum  & 95 / 0       & 5      & 1   & 7.01e6  & 0      & 4.91e4 & 0      & 2  & 0 & 0 & 0 & 108    & 2 \\
\bottomrule
\end{tabular}}
\end{table*}

To answer \ref{rq:colorsaver}, we compare \ColorSaver
against full stack coloring on SPEC CPU2006 INT under QEMU. We
also show that \textbf{\FRESCO completes \texttt{omnetpp} under the SPEC \emph{ref} workload, which exceeds two million live colored allocations, unachievable for \PICASSO alone}\cite{Wang26}.

\Cref{tab:decision-vs-full} reports, per benchmark: escape-analysis classification (Safe/Unsafe), stack segments consumed, split
between calls served by the fast inline path and those taking the slow
end-of-epoch path, maximum call-stack depth, and completed revocation sweeps under the default configuration.  \textbf{\ColorSaver reduces segments needed by an order of magnitude.}
It also keeps instrumentation compact: full coloring grows SPEC CPU2006 binaries by 26.5\%, \ColorSaver by 1.4\%, and the complete configuration including stale-read clearing by 2.0\% (\Cref{tab:spec-code-size}, Appendix~\ref{app:eval}).

\subsection{Performance Evaluation}\label{sec:perfeval}

\begin{figure}[t]
    \centering
    \includegraphics[width=\columnwidth]{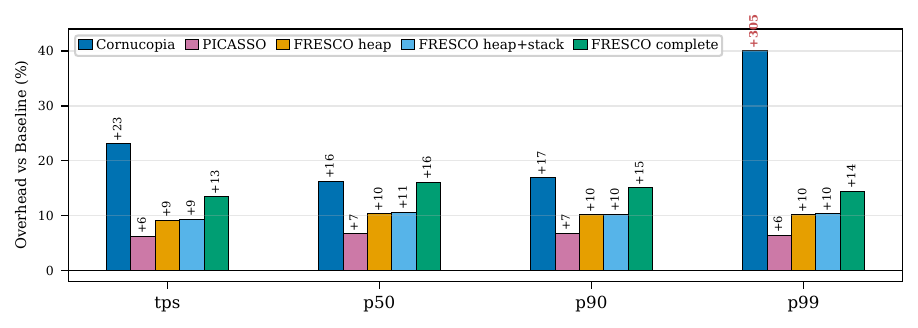}
    \caption{PostgreSQL \texttt{pgbench} latency percentiles and \acrshort{tps}.}\label{fig:cc-pgbench}
\end{figure}
\paragraph{Impact on baseline CHERI-Toooba.} To answer \ref{rq:non-cc-performance}, we analyze \FRESCO's impact on CHERI-Toooba processor logic, energy, and memory. We synthesize \FRESCO CHERI-Toooba for the VCU118 \gls{fpga}. \Cref{tab:hwcost} shows power and resource utilization reported by Xilinx Vivado 2019.1. The overheads over the unmodified CHERI-Toooba are moderate: $\approx$5\% in logic, $\approx$6.40\% in registers, $\approx$8\% in memory. \textbf{\FRESCO incurs minimal hardware-cost compared to \PICASSO}. 

\paragraph{Microbenchmarks.}
To isolate the cost of \stackcoloring, we color \emph{every} function frame in MiBench~\cite{Guthaus01} using a 
freestanding version of \libfresco{} running baremetal on a cycle-accurate Bluesim model of the
\FRESCO softcore. Full coloring adds
a 12.9\% geometric-mean cycle overhead or 7.0\% g.m. excluding the \texttt{limits} microbenchmark, an outlier with 127.7\% overhead representing an extreme worst-case of function calls with almost no computation (\Cref{tab:mibench} in Appendix~\ref{appx:microbenchmarks}).
Each colored call costs 15--25 cycles.
Benchmarks whose hot loop is a function call pay the most---\texttt{bitcount} makes 67501 colored calls through function pointers to single-expression functions for 37.7\% overhead; \texttt{limits} reaches 127.7\%---while
those looping inside a single frame show overheads below measurement noise (\texttt{dijkstra}, \texttt{sha}).

\begin{figure}[t!]
\begin{subfigure}[b]{\columnwidth}
    \centering
    \includegraphics[width=\linewidth]{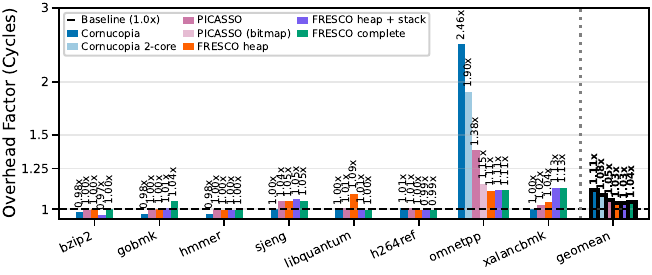}
    \caption{Normalized CPU cycles.}\label{fig:cycles_overhead}
\end{subfigure}
\begin{subfigure}[b]{\columnwidth}
    \centering
    \includegraphics[width=\linewidth]{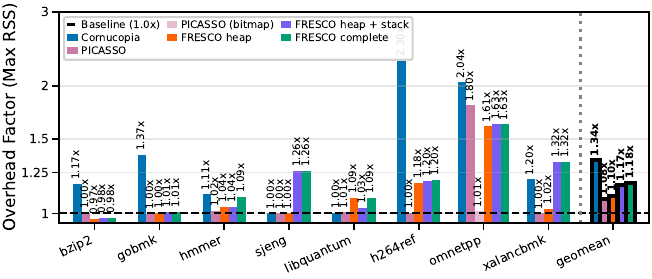}
    \caption{Normalized memory overhead measured as maximum \acrshort{rss}.}\label{fig:memory_overhead}
\end{subfigure}
\caption{SPEC CPU2006 INT results}\label{fig:spec}
\end{figure}

\paragraph{SPEC INT 2006.} To answer \ref{rq:cc-performance}, we evaluate \FRESCO on the PICASSO artifact's setup~\cite{Gulmez26b}: the subset of
SPEC CPU2006 INT with the train input set that runs unmodified under baseline \gls{cheri} on the VCU118 FPGA. \Cref{fig:spec} reports normalized cycle and memory overheads for the CheriBSD allocator on an unmodified CHERI-Toooba core (Cornucopia), for \PICASSO with its original allocator and our bitmap allocator (\Cref{appx:bitmapalloc}); and for \FRESCO in three configurations: segmented heap only (\emph{heap}), segmented heap and stack coloring (\emph{heap+stack}), and heap+stack with stale-read clearing (\emph{complete}). All benchmarks are compiled with optimization level \texttt{-O3}. On omnetpp, the original \PICASSO allocator incurs a 38\% overhead; our bitmap allocator reduces this to 15\%, and the segmented allocator to 11\% with no revocation.
The \emph{stack+heap} and \emph{complete} variants add nothing further, despite 2 stack segments and $9\times10^5$ slow-path colored calls.
The residual cost is not revocation but segment retirement: omnetpp exhausts its segments' colors and takes the retry-allocation path in \texttt{jemalloc} more than 3,000 times, since its caches keep serving memory from retired segments \ifnotabridged before switching to the active one\fi (\Cref{sec:segmentedalloc}). \ifnotabridged We attribute most of the remaining overhead to this path.\fi
xalancbmk performs roughly 460 stack-segment switches across millions of colored calls, yet stays near 13\%, indicating that stack coloring is inexpensive on both the fast and slow paths. 
Overall, \FRESCO incurs $\approx4\%$ \glshyph{gm} run-time overhead, below both Cornucopia ($\approx11\%$) and \PICASSO ($\approx5\%$) despite covering \emph{both} stack and heap.
Memory overhead is $\approx18\%$ \acrshort{gm} (cf. \PICASSO's $\approx8\%$) but the heap-only configuration is only $2\%$ in \PICASSO's favor.
In SPEC, \textbf{\FRESCO closes the stack use-after-return gap with no aggregate run-time cost relative to \PICASSO}.

We also evaluate \FRESCO on SQLite 3.22.0~\cite{sqlite}, PostgreSQL~\cite{postgresql, cheripostgress} and gRPC 1.54.2 \acrshort{qps}~\cite{Tiller25}. The PICASSO artifact builds the benchmarks using \texttt{cheribuild} defaults~\cite{CTSRD26}. We reproduce that setup for comparability to prior work.

\paragraph{SQLite.} SQLite's speedtest1 benchmark exercises different database operations (create, insert, reorder, delete, etc.). \Cref{tab:sqlitephase} lists the phases and \Cref{fig:cc-sqlite-phase} the per-phase results, both in Appendix~\ref{app:eval}.
Overall, \FRESCO (\emph{complete}) incurs a run-time overhead of $\approx10\%$ compared to \PICASSO's $\approx8\%$ and Cornucopia's $\approx11\%$, and a memory overhead of $\approx90\%$, of which $\approx71\%$ relates to the heap; just below Cornucopia's $99\%$.

\paragraph{PostgreSQL \texttt{pgbench}.} 
We run PostgreSQL server on CHERI-Toooba with \texttt{pgbench} on a separate host connected over the local network. We configure a scale factor of 10 and execute $10^3$ transactions with one client. \ifnotabridged To reduce I/O overhead and ensure single-threaded execution suitable for the single-core CHERI-Toooba processor, we\else We \fi disable fsync, synchronous commit, full page writes, statistics collection, and parallel workers. 
\Cref{fig:cc-pgbench} shows Cornucopia's p99 latency rising by over 300\%, a consequence of its revocation sweep in a single-core environment. \FRESCO (\emph{complete}) degrades \gls{tps} by 13.8\%, compared to \PICASSO's 6.2\%, and Cornucopia's 23.1\%: a moderate reduction, but free of Cornucopia's latency peaks at p99. 

\paragraph{gRPC QPS.} The gRPC~\cite{Tiller25} server runs on the \gls{fpga}, and ten clients run on a separate host. Transport security is disabled, each client and server is a single synchronous process, and one channel per client with eight outstanding messages. The clients send 0-byte requests and await responses while recording throughput and latency percentiles.  Each run lasts 120 seconds with a 15‑second warmup, and we report the median of three runs per configuration. \Cref{fig:cc-qps} in Appendix~\ref{app:eval} shows the latency distribution. \FRESCO (\emph{complete}) costs 16.7\% in \gls{qps} against Cornucopia's 15.6\% and \PICASSO's 2.9\%.
The tails separate the designs: at p99, \FRESCO adds 23\% and \PICASSO 5.2\%, while Cornucopia's latency rises $3.7\times$ under revocation. In directly comparable configurations, \FRESCO (\emph{heap}) trails the best-performing \PICASSO by 4.7\% to 2.9\%.

Across all three workloads, the total cost of \FRESCO's complete coverage stays below or comparable to Cornucopia, currently deployed in CheriBSD, which covers only the heap. \textbf{\FRESCO's complete temporal safety therefore costs no more, in aggregate, than today's partial enforcement.}

\section{Related Work}\label{sec:relatedwork}

Stack Temporal safety has been treated as easier to secure~\cite{Dang17}, far rarer than heap-based use-after-free~\cite{Younan15}, and, per Lee et al., \emph{``unlikely to be exploitable''}~\cite{Lee15}; 
a decade of work has accordingly confined run-time mitigation to the heap~\cite{Lee15,Younan15,Dang17,WesleyFilardo20,Filardo24}.
Medium- and high-severity \acrshort{cve}s against stack-allocated state nonetheless continue to be disclosed~\cite{NVD26,NVD26a,NVD26b}.

\paragraph{Compiler-based defenses.}
Clang's \texttt{-Wreturn-stack-address}~\cite{LLVMTeam25a} and GCC's \texttt{-Wreturn-local-addr}~\cite{GCCTeam25} reject only the unconditional error of returning a stack reference directly from its allocating function; references that escape indirectly, through heap or statically allocated memory, evade them. 
Dynamic sanitizers close part of this gap by tracking pointer propagation at run time~\cite{Dang17,LLVMTeam26,GCCTeam26}, but their per-store bookkeeping precludes production deployment.
\FRESCO's enforcement is instead architectural: escape analysis only removes instrumentation the hardware provably does not need.

Clang's \texttt{-ftrivial-auto-var-init} prevents stale reads of capabilities, but writes every byte of every local it applies to; \ColorSaver's stale-read analysis gives the same protection against stale capabilities with fewer stores (\Cref{subsec:uninitialized-read}), but leaves unrelated uninitialized non-capability reads unmasked.

\paragraph{Embedded \gls{cheri} \glspl{cpu}.}
CHERIoT~\cite{Amar23} adapts the \gls{cheri} \gls{isa} to resource-constrained embedded \glspl{cpu} and lets software derive capabilities storable only to stack memory, so stack references cannot reach the heap or a global.
A stack pointer can then outlive its allocation only by being returned in a register or stored through a pointer to a higher frame. The restriction is non-conformant with the C standard, and how much existing code it accommodates is unclear.
\FRESCO targets \gls{cheri} application \glspl{cpu} and leaves stack capabilities free to escape while their frame is live, retaining (\gls{cheri}) C/C++ compatibility.
CheriOS's slinky stack~\cite{Esswood21} builds on SafeStack~\cite{LLVMTeam26a}, 
placing non-escaping locals on a contiguous safe stack and potentially escaping ones on a segmented, monotonically advancing unsafe stack.
Because the unsafe stack pointer never decrements on return, expired memory is not immediately reallocated:
capability bounds keep active functions out of expired frames, and an exhausted segment is quarantined and swept before its physical memory is recycled.

\paragraph{\Gls{cheri} composed with memory versioning.}
Filardo~\cite{Filardo24a}, partially specified in CHERI ISAv9~\cite[Appendix C.5]{Watson23} stamps each allocation and its pointers with a matching version and re-versions the memory on free, so stale capabilities cease to match, closing the reallocation gap without a quarantining.
Versions are held in memory tags whose width is constrained (Arm \glsdesc{mte} exposes 4-bit tags, or 16 versions), so versions must be recycled, and collisions between live and stale pointers degrade the guarantee from deterministic to probabilistic.
Exhaustion falls back to quarantine and Cornucopia-style sweeping, reintroducing the cost versioning was meant to avoid, 
and the composed design has not been implemented.
These shortcomings motivate encoding provenance in the capability itself~\cite{Gulmez26}.

\paragraph{CHERI-D.}
Concurrent to our work, CHERI-D~\cite{Wang26} likewise extends \gls{cheri} with an architecturally-tracked provenance \gls{id} for immediate reuse of freed heap memory.
The designs differ in where the metadata lives and how each lifts \PICASSO's \gls{id} ceiling. 
CHERI-D co-locates an 8-bit \gls{id} with the object data, inline in its cache line or in a per-page \gls{id} table, and checks it against the capability's \gls{id}. A slot thus can be reallocated 254 times before it is quarantined, trading \PICASSO's global color ceiling for a per-slot one.
CHERI-D supports only allocations smaller than a page, leaving larger objects on the sweeping path, and whether allocation \gls{id} in-lining can extend to stack allocations is unclear. 
\FRESCO retains \PICASSO's separately-indexed \gls{pvt} and removes the process-wide color ceiling through \colorsegmentation, so the usable identifier space scales with the number of live segments rather than with the width of a fixed field, and without a per-object reuse bound.

\section{Discussion and Future Work}\label{sec:discussion}

\paragraph{On memory-safety completeness.}
Partial memory safety is not a graceful degradation but a distinct, weaker property whose \emph{residual gaps} remain individually exploitable.
Conventional engineering has failed to eliminate memory-safety vulnerabilities because a single error in a multi-million-line codebase can yield total control to \gls{adv}; under this economy of failure, a mechanism that enforces a property for every allocation class but one leaves an intact attack surface rather than a proportionally reduced guarantee.
Watson et al.~\cite{Watson25} accordingly argue that protection technologies should be classified not by attack class but by their contribution to memory-safety properties such as spatial and temporal safety, and by whether that contribution is \emph{complete}.

\ifnotabridged
\acrshort{etsi}'s  draft \glsdesc{ts} on memory-safety requirements~\cite{ETSI26}, made available for public review in July 2026, adopts this framing: completeness limitations denote incomplete enforcement of a property otherwise present in a system's semantics~\cite[clause 6.3]{ETSI26}, and \gls{cheri} C/C++ appears there as an example of a system which protects mappings, heap allocations, globals, and thread-local storage, \textbf{but not stack allocations}, hence implementing temporal safety incompletely.
\FRESCO closes precisely the gap an emerging standardization framework already documents, extending temporal safety to stack use-after-return alongside heap protection on application-class processors.
\fi

Recall from \Cref{sec:system-and-adversary-model} that \FRESCO's completeness claim is one of \emph{escape routes}, not of allocation granularity. Intra-frame lifetime violations, in which a reference outlives the lexical scope of its referent while the allocating activation remains live (use-after-scope), are not addressed: coloring is assigned per activation, so all allocations within a frame are indistinguishable by provenance, and, being within the bounds of a live object, such accesses are equally invisible to \gls{cheri}'s spatial checks.
This is a granularity decision rather than a limitation of \stackcoloring. Colors can be assigned per scope rather than per frame at the cost of color pressure. However, whereas frame teardown is an architecturally visible event that a prologue/epilogue can instrument, block-lifetime boundaries exist only in the compiler's \gls{ir} and have no run-time manifestation unless one is manufactured~\cite{Nyman19}.

\paragraph{Relevancy for \gls{cheri} standardization.}
RISC-V International's \gls{cheri} Special Interest and Task Groups are driving CHERI-RISC-V toward ratification~\cite{Kurd26}, and commercial silicon and softcores are already emerging around the CHERIoT and Codasip ecosystems.
The draft defines \texttt{RVY}, a family of base \glspl{isa} adding \gls{cheri} to 64-bit RISC-V, together with optional extensions~\cite{Aird25,Aird25a} enabling temporal-safety enforcement analogous to Cornucopia Reloaded.
\ifnotabridged
The \emph{Svucrg} extension~\cite{Aird25} adds a \gls{crg} bit to each \gls{pte} and a \gls{ucrg} bit in \gls{cpu} status registers, mirroring Cornucopia Reloaded's per-\gls{pte} revocation epochs: capability loads succeed only while \gls{crg} equals \gls{ucrg}, and otherwise raise a \gls{cheri} load-page-fault.
Toggling \gls{ucrg} therefore invalidates stale capabilities across all pages at once, removing the pass that must otherwise pre-mark every \gls{pte} unreadable at the start of a concurrent sweep.
The \emph{Svy} extension's \gls{cw} bit~\cite{Aird25a} governs whether a page may hold capabilities and further cuts sweep cost: a page with \gls{cw} unset cannot hold valid capabilities, so it can be skipped entirely and loads from it clear the validity tag.
\fi
An earlier concern that scarce free \gls{pte} bits would force a simplified encoding, and with it a revocation state machine slower than \gls{cheri} \gls{isa}v9's~\cite{Gulmez26}, has not materialized: the draft under review uses five \gls{pte} bits, mirroring \gls{isa}v9~\cite[4.3.12]{Watson23a}.

\FRESCO{} complements these efforts rather than competing with them.
Building on \ccs, it reduces how \emph{often} revocation must run rather than how the sweep is realized, and so composes with \gls{rv64y}.
Its provenance indirection rests on modest ISA-level additions (\gls{pvtr}, \gls{ccsettype}, \gls{pvb} check, segmentation controls), is core-agnostic, and, through \colorsegmentation, removes the dependence on a \ccpid{} of a fixed width.
We prototype on the \gls{cheri} \gls{isa}v9 CHERI-Toooba core, the most mature openly available multi-stage application core for \gls{cheri}, and expect \FRESCO{} to apply to future large-pipeline CHERI processors; adaptation to \gls{rv64y} is left as future work.

\ifnotabridged
\input{sections/conclusion}
\fi
\ifnotarxiv
\input{sections/ethicalconsiderations}
\input{sections/openscience}
\fi
\ifnotanonymous
\filbreak
\section*{Acknowledgments}

We thank our colleague at Ericsson: Håkan Englund, for his feedback on this manuscript and the discussions on early designs, leading up to this work.
This work is supported in part by the Wallenberg Visiting Professor Program, the Natural Sciences and Engineering Research Council of Canada (grant number RGPIN-2026- 04826).
\fi

\bibliographystyle{IEEEtran}
\bibliography{main, local}

\appendix

\section{\PICASSO \gls{unr} allocator}\label{appx:unralloc}

The \gls{unr} allocator is \PICASSO's adaptation of FreeBSD's unit-number allocator (\texttt{alloc\_unr}), a facility allocating integer \glspl{id} that \PICASSO uses to assign and reclaim \ccpids (colors).
It stores the claimed/available state of all \glspl{id} in compressed form as a doubly-linked list whose nodes are either \emph{runs}, which compactly encode consecutive ranges of allocated or free \glspl{id}, or \emph{bitmaps}, which cover fragmented regions. Bitmaps are a fixed 512 bits, and they are formed only where they replace at least three runs.

To allocate a color, the \gls{unr} traverses the list until it finds an available \gls{id}.
While the color space remains unfragmented, a free run is always adjacent to the allocated run, so the allocator simply decrements the free run's length, increments the allocated run's length, and returns the next sequential \gls{id}.
If an adjacent node is instead a bitmap, which becomes more likely as color fragmentation increases, the allocator must search for the first free bit, set it, and return the corresponding \gls{id}.

\begin{figure*}[t!]
\begin{subfigure}[b]{0.45\linewidth}
    \centering
    \includegraphics[width=\linewidth]{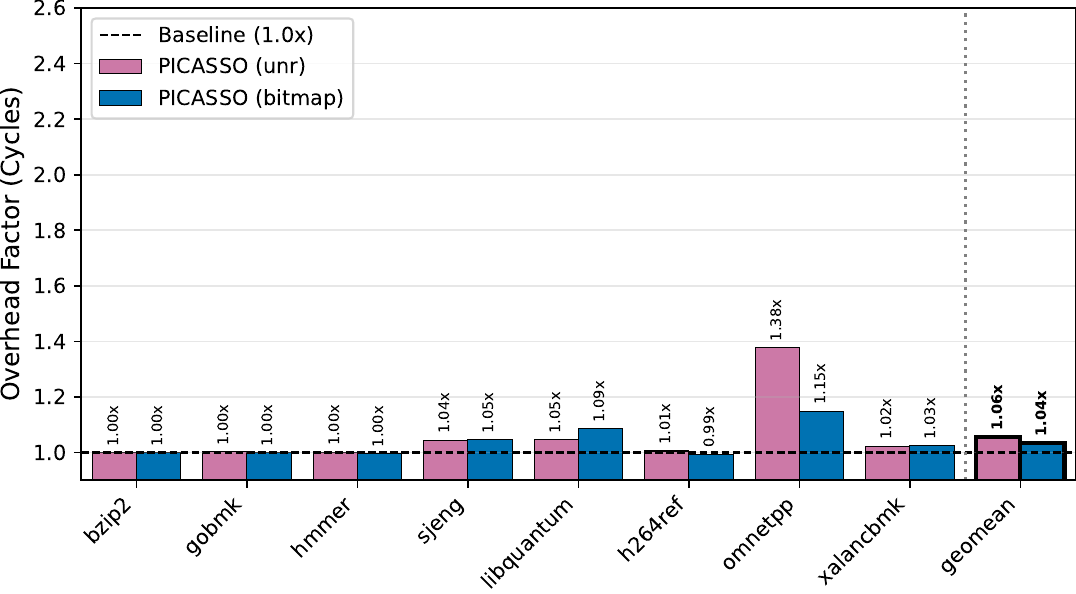}
    \caption{Normalized CPU cycles.}
    \label{fig:cycles_overhead}
\end{subfigure}
\hfill
\begin{subfigure}[b]{0.45\linewidth}
    \centering
    \includegraphics[width=\linewidth]{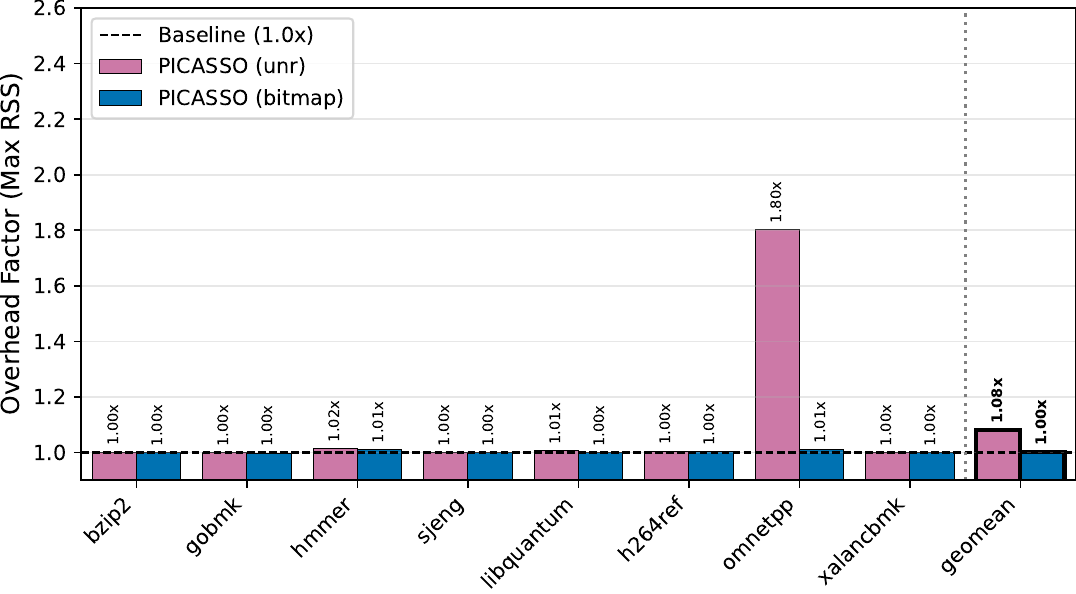}
    \caption{Normalized memory overhead measured as maximum \acrshort{rss}.}
    \label{fig:memory_overhead}
\end{subfigure}
\caption{Comparison between \gls{unr} and bitmap allocator.}
\end{figure*}

Reclamation first locates the node holding the \gls{id}.
In a bitmap, the corresponding bit is cleared; in a run, the run is split into three parts whose middle part is the reclaimed \gls{id}, after which a \textit{collapse\_unr} function merges neighboring runs and bitmaps to mark the \gls{id} available.
In both cases, adjacent nodes may be merged or converted between representations to keep the structure compact.

FreeBSD's \gls{unr} provides no bulk reclamation of \glspl{id}.
Because \PICASSO frees a large number of \glspl{id} at each revocation, calling the per-\gls{id} free function in a loop would be prohibitively inefficient.
\PICASSO therefore adds a \texttt{free\_many\_unr()} function that rebuilds the internal \gls{unr} structure from scratch:
it allocates a new \gls{unr} header, iterates through the revoker's \gls{pvt} snapshot to determine which \glspl{id} remain in use, reconstructs the full \gls{unr} structure, and then frees the old structure. 

This solves the bulk free problem, but at a significant cost in memory and run-time. The old and new data structures coexist in memory for the duration of the \texttt{free\_many\_unr()} call, the rebuild performs a large number of node allocations, and subsequent calls still perform structural maintenance to recompact the result.
These costs are inherent to the \gls{unr} design: freeing a single \gls{id} in the middle of a run splits the run into three nodes and forces restructuring, so bulk-freeing \glspl{id} scattered across the structure requires either modifying every affected node individually, or discarding and rebuilding the structure wholesale. The \gls{unr} allocator's data structure is therefore ill suited for efficient in-place bulk updates.

\subsection{Bitmap-based \ccpid allocation.}\label{appx:bitmapalloc}
Scaling the color space through \colorsegmentation makes the cost of \gls{id} bookkeeping more, not less, important.
\PICASSO's residual memory overhead (up to $\approx80\%$ during revocation on omnetpp) stems from fragmentation of the color space in its \gls{unr} allocator (\Cref{sec:temporal-safety-gaps}).
\FRESCO replaces the \gls{unr} allocator with a flat bitmap that represents each identifier by a single bit.
This yields four characteristics the \gls{unr} lacks:
\begin{inparaenum}[1)]
  \item Its memory footprint is constant: $n=256$ KiB for a $2^{21}$-identifier namespace, irrespective of allocation pattern, fragmentation, or revocation count.
  \item No operation allocates dynamically so allocation and free require no system calls; the \gls{unr} allocator, by contrast, frequently allocates internal nodes, particularly once revocation has fragmented the structure.
  \item The bitmap representation is independent of fragmentation and requires no structural maintenance, whereas the \gls{unr} allocator performs frequent splitting, merging, and representation conversion.
  \item Because the allocator bitmap and \gls{pvt} supplied at revocation have identical layout, bulk reclamation reduces to word-granular AND-NOT operations over [$n/64$] words.
\end{inparaenum}

\paragraph{Design and implementation.}
The allocator stores the bitmap in a statically sized array and uses FreeBSD's bitstring primitives for scanning and bulk bit manipulation. As in the \gls{unr} allocator, `\texttt{low}` and `\texttt{high}` delimit the allocatable \gls{id} range, `num\_bits` records the number of identifiers in that range, `busy` the number of currently allocated \glspl{id}, `hint` the position at which the next search begins, and `lock` a spinlock ensuring thread safety across allocation and free. 

Allocation uses \texttt{hint} as a fast path: if the \texttt{hint} points to a free bit, it is claimed without any scan. Otherwise the allocator scans from \texttt{hint} to \texttt{num\_bits} and, failing that, wraps around to scan from $0$ up to but excluding hint. On success, \texttt{hint} advances to one past the claimed position, and the returned identifier is the bit position plus \texttt{low}, mapping the internal bitmap offset to the hardware-configured color range.

Bulk freeing aligns the allocator's bitmap with the sealing bitmap supplied by the revoker and, for each 64-bit word, ANDs it with the negation of the corresponding sealing bitmap word. It then updates \texttt{busy} and sets \texttt{hint} to the beginning of the first word a bit was freed from. The bulk freeing is a single pass over the bitmap and requires no allocation.

\subsection{Benchmark comparison against \gls{unr}}
We evaluate the bitmap allocator using the same setup as in Gülmez et al.~\cite{Gulmez26}: the subset of SPEC CPU2006 INT benchmarks that run unmodified on the baseline CHERI system, with the train input set on the VCU118 FPGA. We compare \PICASSO using our bitmap allocator against \PICASSO using its original \gls{unr} allocator.

\Cref{fig:cycles_overhead} reports normalized CPU cycles. On the seven benchmarks that never trigger revocation, the two allocators perform equivalently, with the remaining differences within measurement noise. On omnetpp, the only benchmark triggering revocations, the bitmap allocator reduces the cycle overhead by $23\%$ compared to \PICASSO's \gls{unr}.

\Cref{fig:memory_overhead} reports normalized memory overhead as maximum \gls{rss}. The bitmap allocator's footprint is a constant $\approx 256$KiB irrespective of workload characteristics. The effect is clearest on omnetpp, where the bitmap allocator stays at 1.01x memory usage, while \gls{unr} peaks at 1.80x.

\section{Escape Analysis}
\label{app:escape-details}

\subsection{LLVM Facts}\label{tab:escape-fact-extraction}

LLVM uses static single assignment form: a function argument and each
instruction result is an LLVM \emph{value}.  The analysis records only values
that may carry a capability. We write $f$ and $g$ for functions, $c$ for a call
instruction, $i$ for an argument position, and $v,d,s,p$ for LLVM values. The
following facts are extracted from LLVM IR.

\begin{compactitem}
  \item $\mathsf{Copy}(f,d,s)$ represents a capability-preserving SSA
        operation from source $s$ to destination $d$ in $f$.  Such operations
        include casts, \texttt{getelementptr}, \texttt{phi}, \texttt{select},
        and aggregate or vector extraction and insertion.
  \item $\mathsf{Load}(f,d,p)$ represents an LLVM load through pointer $p$
        whose result is $d$.
  \item $\mathsf{Store}(f,p,s)$ represents an LLVM store of $s$ through
        pointer $p$.
  \item $\mathsf{Parameter}(f,i,v)$ says that $v$ is formal parameter $i$ of
        $f$.  A \emph{formal parameter} is the value received by the callee.
  \item $\mathsf{ReturnValue}(f,v)$ says that an LLVM \texttt{ret} in $f$
        returns $v$.
  \item $\mathsf{Call}(f,c)$ says that call instruction $c$ occurs in $f$.
        $\mathsf{CallArgument}(c,i,v)$ records its actual argument $i$, where
        an \emph{actual argument} is the value supplied by the caller.
        $\mathsf{CallResult}(c,v)$ records the call result.
  \item $\mathsf{InlineCall}(c,g)$ says that call $c$ may invoke $g$, so the
        analysis connects the actual arguments at $c$ to the formal parameters
        of $g$.
        $\mathsf{ReturnCall}(c,g)$ says that values returned by $g$ may flow to
        the result of $c$.
  \item $\mathsf{RetainValue}(f,v)$ says that $v$ may be retained after $f$
        returns, for example by persistent memory or an unresolved call.
  \item $\mathsf{ObserveValue}(f,v)$ says that code outside the analysis may
        access memory through $v$ but may not retain $v$ itself.  This models,
        for example, an LLVM \texttt{nocapture} argument.
  \item $\mathsf{FrameOrigin}(f,v)$ says that $v$ may point into the current
        stack frame.  LLVM \texttt{alloca} results and instructions that
        expose the stack pointer are frame origins.
\end{compactitem}

\subsection{Algorithm}

One execution of a function is an \emph{activation}, and its stack storage is
its \emph{frame}.  The frame escapes if a capability pointing into it remains
reachable after the activation returns.  This can happen in two ways: the
capability can be stored in persistent memory, or it can be returned.  The
analysis therefore performs one \emph{Retain} search and one \emph{Return}
search.  Both searches start at their respective escape endpoints, propagate
through calls and LLVM value flow, and report a function when they reach one
of its frame origins.

Loads and stores require more than SSA reachability.  The analysis attaches a
short \emph{access path} to each value.  An access path is a record of the
memory edges that must be crossed to reach the capability. Formally, a path is a
word $\ell\in P=\{L,R\}^{\leq 5}$, where $\ell=\epsilon$ means that the
capability is the value itself.  $L$ labels the content edge of an abstract
memory cell, and $R$ labels the edge for the value supplied to a store.

The four path operations describe how a load or store changes this record:

\begin{compactitem}
  \item $\mathsf{prepend}_{L}(\ell)$ adds a leading $L$.  When a reachable
        value is stored through pointer $p$, this records that the value can
        subsequently be found in the content of $p$.
  \item $\mathsf{prepend}_{LR}(\ell)$ returns the set
        $\{\mathsf{prepend}_{L}(\ell),\mathsf{prepend}_{R}(\ell)\}$, namely
        both $L\ell$ and $R\ell$; here $\mathsf{prepend}_{R}$ symmetrically
        adds $R$.  When reasoning backward from a load result
        to its pointer, the analysis conservatively considers both sides of
        the abstract memory cell.
  \item $\mathsf{take}_{L}(\ell)$ removes a leading $L$, if present.  Thus a
        path $L\ell$ at a load pointer becomes path $\ell$ at the loaded
        value.
  \item $\mathsf{take}_{R}(\ell)$ removes a leading $R$, if present.  Thus a
        path $R\ell$ at a store pointer becomes path $\ell$ at the stored
        value.
\end{compactitem}

Whenever a rule writes $\ell'\in S$, it means that the rule is instantiated
once for every element $\ell'$ of the set $S$; it is not an equality between a
path and a set.  For example, if a pointer has path $L R\epsilon$, a load through that pointer
uses $\mathsf{take}_{L}$ and leaves path $R\epsilon$ on the loaded value.  If
the pointer has path $R\epsilon$, a store through it uses
$\mathsf{take}_{R}$ and leaves $\epsilon$ on the stored value.  A path that
does not start with the required letter produces no result.  Paths are capped
at length five; at that limit the rules keep conservative suffix variants
rather than dropping the flow.  Finally,
$P^+=P\setminus\{\epsilon\}$ is the set of nonempty paths.

Each search has two propagation relations.  \textsf{RetainUsed} and
\textsf{ReturnUsed} record ordinary reachability for the corresponding
search.  \textsf{RetainDerived} and \textsf{ReturnDerived} mark the subset of
that reachability which may also be followed backward through the definition
of a value.  Every Derived fact also has a corresponding Used fact.

The following named abbreviations make the call rules readable:
\[
\begin{aligned}
\mathsf{ArgFlow}(f,c,g,i,a,p)\equiv{}&
  \mathsf{Call}(f,c)\\
  &{}\land\mathsf{InlineCall}(c,g)\\
  &{}\land\mathsf{CallArgument}(c,i,a)\\
  &{}\land\mathsf{Parameter}(g,i,p),\\
\mathsf{RetFlow}(f,c,g,x,r)\equiv{}&
  \mathsf{Call}(f,c)\\
  &{}\land\mathsf{ReturnCall}(c,g)\\
  &{}\land\mathsf{CallResult}(c,x)\\
  &{}\land\mathsf{ReturnValue}(g,r).
\end{aligned}
\]
Thus, $\mathsf{ArgFlow}$ connects an actual argument $a$ to the matching formal
$p$, and $\mathsf{RetFlow}$ connects a returned value $r$ to the call result
$x$.

\subsection{Rules}

The rules in this subsection are the rules covered by the Rocq proof.  In the
common propagation rules, \textsf{Used} and \textsf{Derived} are shorthand: read
each rule once with \textsf{RetainUsed}/\textsf{RetainDerived} and once with
\textsf{ReturnUsed}/\textsf{ReturnDerived}.

\paragraph{Endpoints that retain.}
\begin{compactitem}
  \item $\mathsf{RetainUsed}(f,v,\ell)\leftarrow
        \mathsf{RetainValue}(f,v),\ell\in P$.\\
        This rule adds a $\mathsf{RetainUsed}$ fact for every path in $P$.
  \item $\mathsf{RetainDerived}(f,v,\ell)\leftarrow
        \mathsf{RetainValue}(f,v),\ell\in P$.\\
        This rule adds a $\mathsf{RetainDerived}$ fact for every path in $P$.
  \item $\mathsf{RetainUsed}(f,v,\ell)\leftarrow
        \mathsf{ObserveValue}(f,v),\ell\in P^+$.\\
        This rule adds $\mathsf{RetainUsed}$ for every nonempty path in $P^+$.
  \item $\mathsf{RetainDerived}(f,v,\ell)\leftarrow
        \mathsf{ObserveValue}(f,v),\ell\in P^+$.\\
        This rule adds $\mathsf{RetainDerived}$ for every nonempty path in
        $P^+$.
  \item $\mathsf{RetainUsed}(f,p,\ell)\leftarrow
        \mathsf{Parameter}(f,i,p),\ell\in P^+$.\\
        This rule adds $\mathsf{RetainUsed}$ for every nonempty path at formal
        $p$.
  \item $\mathsf{RetainDerived}(f,p,\ell)\leftarrow
        \mathsf{Parameter}(f,i,p),\ell\in P^+$.\\
        This rule adds $\mathsf{RetainDerived}$ for every nonempty path at
        formal $p$.
\end{compactitem}

\paragraph{Calls.}
\begin{compactitem}
  \item $\mathsf{RetainUsed}(f,a,\ell)\leftarrow\allowbreak
        \mathsf{ArgFlow}(f,c,g,i,a,p),\allowbreak
        \mathsf{RetainUsed}(g,p,\ell)$.\\
        This rule copies $\mathsf{RetainUsed}$ from formal $p$ in $g$ to actual
        $a$ in $f$.
  \item $\mathsf{Used}(g,p,\ell)\leftarrow\allowbreak
        \mathsf{ArgFlow}(f,c,g,i,a,p),\allowbreak
        \mathsf{Used}(f,a,\ell)$.\\
        This rule copies $\mathsf{Used}$ from actual $a$ in $f$ to formal $p$ in
        $g$.
  \item $\mathsf{Used}(f,a,\ell)\leftarrow\allowbreak
        \mathsf{ArgFlow}(f,c,g,i,a,p),\allowbreak
        \mathsf{Derived}(g,p,\ell)$.\\
        This rule copies $\mathsf{Derived}$ from formal $p$ in $g$ to
        $\mathsf{Used}$ at actual $a$ in $f$.
  \item $\mathsf{Derived}(f,a,\ell)\leftarrow\allowbreak
        \mathsf{ArgFlow}(f,c,g,i,a,p),\allowbreak
        \mathsf{Derived}(g,p,\ell)$.\\
        This rule copies $\mathsf{Derived}$ from formal $p$ in $g$ to
        $\mathsf{Derived}$ at actual $a$ in $f$.
  \item $\mathsf{Used}(f,x,\ell)\leftarrow\allowbreak
        \mathsf{RetFlow}(f,c,g,x,r),\allowbreak
        \mathsf{Used}(g,r,\ell)$.\\
        This rule copies $\mathsf{Used}$ from returned value $r$ in $g$ to call
        result $x$ in $f$.
  \item $\mathsf{Used}(g,r,\ell)\leftarrow\allowbreak
        \mathsf{RetFlow}(f,c,g,x,r),\allowbreak
        \mathsf{Derived}(f,x,\ell)$.\\
        This rule copies $\mathsf{Derived}$ from call result $x$ in $f$ to
        $\mathsf{Used}$ at returned value $r$ in $g$.
  \item $\mathsf{Derived}(g,r,\ell)\leftarrow\allowbreak
        \mathsf{RetFlow}(f,c,g,x,r),\allowbreak
        \mathsf{Derived}(f,x,\ell)$.\\
        This rule copies $\mathsf{Derived}$ from call result $x$ in $f$ to
        $\mathsf{Derived}$ at returned value $r$ in $g$.
\end{compactitem}

\paragraph{Endpoints that return.}
\begin{compactitem}
  \item $\mathsf{ReturnUsed}(f,r,\ell)\leftarrow
        \mathsf{ReturnValue}(f,r),\ell\in P$.\\
        This rule adds a $\mathsf{ReturnUsed}$ fact for every path in $P$.
  \item $\mathsf{ReturnDerived}(f,r,\ell)\leftarrow
        \mathsf{ReturnValue}(f,r),\ell\in P$.\\
        This rule adds a $\mathsf{ReturnDerived}$ fact for every path in $P$.
\end{compactitem}

\paragraph{Copies.}
\begin{compactitem}
  \item $\mathsf{Used}(f,d,\ell)\leftarrow
        \mathsf{Copy}(f,d,s),\mathsf{Used}(f,s,\ell)$.\\
        This rule copies $\mathsf{Used}$ from source $s$ to destination $d$.
  \item $\mathsf{Used}(f,s,\ell)\leftarrow
        \mathsf{Copy}(f,d,s),\mathsf{Derived}(f,d,\ell)$.\\
        This rule copies $\mathsf{Derived}$ at destination $d$ to
        $\mathsf{Used}$ at source $s$.
  \item $\mathsf{Derived}(f,s,\ell)\leftarrow
        \mathsf{Copy}(f,d,s),\mathsf{Derived}(f,d,\ell)$.\\
        This rule copies $\mathsf{Derived}$ at destination $d$ to
        $\mathsf{Derived}$ at source $s$.
\end{compactitem}

\paragraph{Loads and stores.}
\begin{compactitem}
  \item $\mathsf{Used}(f,p,\ell')\leftarrow
        \mathsf{Load}(f,d,p),\mathsf{Derived}(f,d,\ell),
        \ell'\in\mathsf{prepend}_{LR}(\ell)$.\\
        For each $\ell'\in\mathsf{prepend}_{LR}(\ell)$, this rule copies
        $\mathsf{Derived}$ at load result $d$ to $\mathsf{Used}$ at pointer $p$.
  \item $\mathsf{Derived}(f,p,\ell')\leftarrow
        \mathsf{Load}(f,d,p),\mathsf{Derived}(f,d,\ell),
        \ell'\in\mathsf{prepend}_{LR}(\ell)$.\\
        For each $\ell'\in\mathsf{prepend}_{LR}(\ell)$, this rule copies
        $\mathsf{Derived}$ at load result $d$ to $\mathsf{Derived}$ at pointer
        $p$.
  \item $\mathsf{Used}(f,d,\ell')\leftarrow
        \mathsf{Load}(f,d,p),\mathsf{Used}(f,p,\ell),
        \ell'\in\mathsf{take}_{L}(\ell)$.\\
        For each $\ell'\in\mathsf{take}_{L}(\ell)$, this rule copies
        $\mathsf{Used}$ at pointer $p$ to loaded value $d$.
  \item $\mathsf{Used}(f,s,\ell')\leftarrow
        \mathsf{Store}(f,p,s),\mathsf{Used}(f,p,\ell),
        \ell'\in\mathsf{take}_{R}(\ell)$.\\
        For each $\ell'\in\mathsf{take}_{R}(\ell)$, this rule copies
        $\mathsf{Used}$ at pointer $p$ to stored value $s$.
  \item $\mathsf{Derived}(f,s,\ell')\leftarrow
        \mathsf{Store}(f,p,s),\mathsf{Used}(f,p,\ell),
        \ell'\in\mathsf{take}_{R}(\ell)$.\\
        For each $\ell'\in\mathsf{take}_{R}(\ell)$, this rule copies
        $\mathsf{Used}$ at pointer $p$ to $\mathsf{Derived}$ at stored value
        $s$.
  \item $\mathsf{Used}(f,p,\mathsf{prepend}_{L}(\ell))\leftarrow
        \mathsf{Store}(f,p,s),\mathsf{Used}(f,s,\ell)$.\\
        This rule prefixes $L$ to the path at stored value $s$ and adds the
        resulting $\mathsf{Used}$ fact at pointer $p$.
  \item $\mathsf{Derived}(f,p,\mathsf{prepend}_{L}(\ell))\leftarrow
        \mathsf{Store}(f,p,s),\mathsf{Used}(f,s,\ell)$.\\
        This rule prefixes $L$ to the path at stored value $s$ and adds the
        resulting $\mathsf{Derived}$ fact at pointer $p$.
\end{compactitem}

\paragraph{Decision.}
\begin{compactitem}
  \item $\mathsf{UnsafeFunction}(f)\leftarrow
        \mathsf{FrameOrigin}(f,o),
        \mathsf{RetainUsed}(f,o,\epsilon)$.\\
        This rule derives $\mathsf{UnsafeFunction}(f)$ from a root
        $\mathsf{RetainUsed}$ fact at a frame origin.
  \item $\mathsf{UnsafeFunction}(f)\leftarrow
        \mathsf{FrameOrigin}(f,o),
        \mathsf{ReturnUsed}(f,o,\epsilon)$.\\
        This rule derives $\mathsf{UnsafeFunction}(f)$ from a root
        $\mathsf{ReturnUsed}$ fact at a frame origin.
\end{compactitem}

\subsection{Rocq Guarantee}\label{app:escape-proof}

The Rocq theorem proves that the basic algorithm has no false negatives for
the abstract machine used by the proof.  More precisely, suppose the input
facts cover every parameter, persistent endpoint, copy, load, store, call, and
return in a normalized program.  If an activation returns while one of its
stack allocations is still reachable from a returned capability, persistent
memory, or a value in a still-live activation, then the rules derive
$\mathsf{UnsafeFunction}$ for that activation's function.  Equivalently, if
the rules classify a function as safe, none of these post-return paths exists
in the abstract machine.  The theorem permits false positives and assumes
that LLVM fact extraction is complete.

\subsection{Optimization}\label{app:escape-optimization}

The production analysis uses three refinements around the basic algorithm.
The theorem above covers only the basic rules.  The refinements are summarized
below.

\begin{compactitem}
  \item \emph{Function summaries.}  For each function, the analysis records
        whether a returned capability may be derived from a particular formal
        parameter or from the caller's frame.  A call can then apply this
        summary directly to its actual argument and result instead of
        repeatedly expanding the callee's return flow.
  \item \emph{Shared-value propagation.}  Globals, aliases, and constant
        expressions are shared by many functions.  The analysis computes
        their capability flow once, then imports only the portion reachable
        from a shared value mentioned by each function.  This avoids building
        one duplicated shared graph per function.
  \item \emph{Variadic caller-frame modeling.}  A variadic callee can obtain a
        capability to its caller's argument area through \texttt{va\_start}.
        The analysis models this as a hidden formal parameter and summarizes
        whether the callee retains or returns it.  Calls then transfer that
        effect to the caller's frame.
\end{compactitem}

\section{Stale-Read Analysis Facts and Rules}
\label{app:freshness-details}

A CHERI memory \emph{tag} records whether a memory word contains a valid
capability.  A \emph{stale capability} is a tagged capability left by an older
use of the same stack storage.  The compiler initially inserts a clearing
store for each source variable that might expose such a capability.  A
\emph{candidate} is one such clear.  A candidate's \emph{root} is an LLVM value
associated with the storage protected by that clear; one candidate may have
several roots after optimization.

The analysis uses these input facts:
\begin{compactitem}
  \item $\mathsf{Candidate}(k,v)$: value $v$ is a root of candidate $k$.
  \item $\mathsf{MustRetain}(k)$: incomplete or unsupported information forces
        candidate $k$ to keep its clear.
  \item $\mathsf{AddressCopy}(d,s)$: $d$ preserves the address represented by
        $s$.  Casts, address calculations, selections, and merges produce this
        fact.
  \item $\mathsf{NonStackOrigin}(v)$: $v$ comes from a parameter, global,
        constant, or function value rather than the local stack variable whose
        clear is being decided.
  \item $\mathsf{TagSink}(v)$: using $v$ may load, expose, or publish a memory
        tag.  A \emph{sink} is simply such an endpoint.
  \item $\mathsf{ClosedTarget}(c,g)$: the extractor knows the complete set of
        functions that call site $c$ may invoke, and $g$ is one of them.
  \item $\mathsf{CallArgument}(c,i,a)$ and $\mathsf{Parameter}(g,i,p)$:
        actual $a$ is passed to formal $p$ at position $i$.
\end{compactitem}

$\mathsf{MayObserveTag}(v)$ is derived when $v$ may reach a tag sink.
$\mathsf{RetainClear}(k)$ is the final decision to keep candidate $k$'s clear.
The complete Datalog program has eight rules:

\begin{enumerate}[label=\textbf{S\arabic*.},leftmargin=*,itemsep=2pt]
  \item
  $\mathsf{MayObserveTag}(v)\leftarrow\mathsf{TagSink}(v)$.
  Traversal starts at every sink and works back toward candidate roots.

  \item
  $\mathsf{MayObserveTag}(s)\leftarrow
  \mathsf{AddressCopy}(d,s),\mathsf{MayObserveTag}(d)$.
  If a derived address may reach a sink, its source may also reach it.

  \item
  $\mathsf{MayObserveTag}(d)\leftarrow
  \mathsf{AddressCopy}(d,s),\mathsf{MayObserveTag}(s)$.
  The reverse direction conservatively treats address copies as possible
  aliases; an \emph{alias} is another value that may name the same storage.

  \item
  $\mathsf{NonStackOrigin}(d)\leftarrow
  \mathsf{AddressCopy}(d,s),\mathsf{NonStackOrigin}(s)$.
  A derived address keeps the source's non-stack origin.

  \item
  $\mathsf{MayObserveTag}(a)\leftarrow
  \mathsf{ClosedTarget}(c,g),\mathsf{CallArgument}(c,i,a), \\
  \mathsf{Parameter}(g,i,p),\mathsf{MayObserveTag}(p)$.
  If a known callee may observe a formal, the matching actual may be observed.

  \item
  $\mathsf{RetainClear}(k)\leftarrow
  \mathsf{Candidate}(k,v),\mathsf{MayObserveTag}(v)$.
  Keep the clear when any root of the candidate may reach a sink.

  \item
  $\mathsf{RetainClear}(k)\leftarrow
  \mathsf{Candidate}(k,v),\mathsf{NonStackOrigin}(v)$.
  Keep the clear when an optimized root is not proven to be source-owned stack
  storage.

  \item
  $\mathsf{RetainClear}(k)\leftarrow\mathsf{MustRetain}(k)$.
  Keep the clear whenever extraction explicitly requests the conservative
  fallback.
\end{enumerate}

\newpage
\onecolumn
\section{Segment-aware \texttt{malloc()} and \texttt{free}}\label{appx:mallocfree-figs}
\begin{figure*}[h]
\begin{subfigure}[b]{0.5\textwidth}
\centering
    \includegraphics[width=0.8\columnwidth]{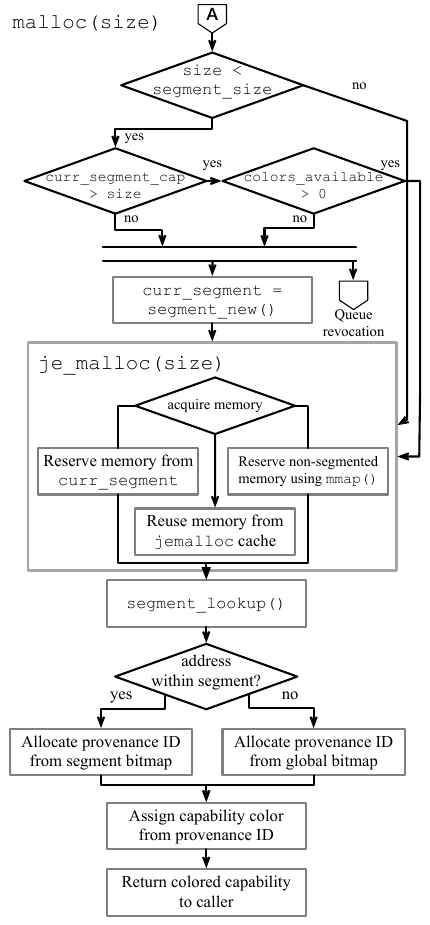}
    \caption{Flowchart of segmented \texttt{malloc()} operation.}\label{fig:segmentedmalloc}
\end{subfigure}
\begin{subfigure}[b]{0.5\textwidth}
\centering
    \includegraphics[width=0.8\columnwidth]{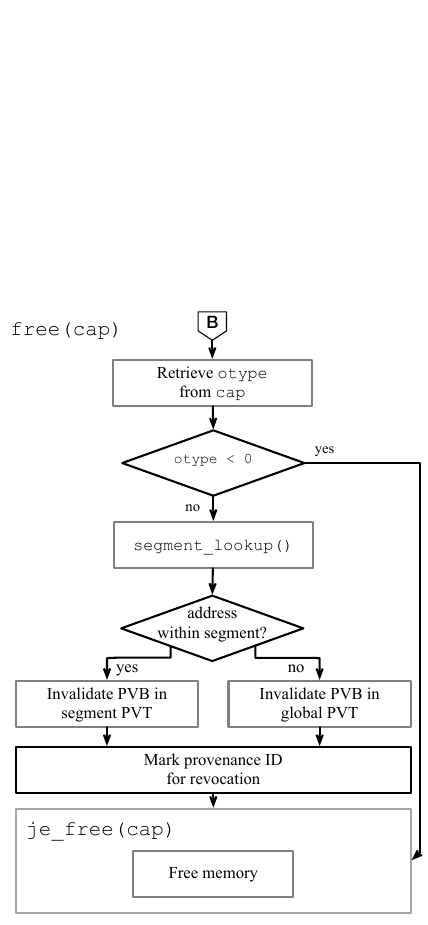}
    \caption{Flowchart of segmented \texttt{free()} operation.}\label{fig:segmentedfree}
\end{subfigure}
\caption{Illustrations of segmented \texttt{malloc()} and \texttt{free()} from \Cref{sec:segmentedalloc}.}
\end{figure*}

\newpage
\twocolumn
\section{Supplementary Evaluation}\label{appx:microbenchmarks}\label{app:eval}

\begin{table}[h]
\centering
\small
\caption{FRESCO stack-coloring overhead on MiBench (\PICASSO, \FRESCO with Full coloring).
The cost is fixed per colored call ($\approx$15--25 cycles)}\label{tab:mibench-fresco}
\begin{tabular}{lrrrr}
\toprule
Benchmark & \multicolumn{2}{c}{Cycles} & Overhead & Colored \\
\cmidrule(lr){2-3}
 & Baseline & Colored & & calls \\
\midrule
adpcm\_decode &   3\,343\,861 & 3\,407\,442 & $+1.9\,\%$   &    801 \\
adpcm\_encode &   2\,998\,361 & 3\,025\,730 & $+0.9\,\%$   &    718 \\
aes           &      72\,906 &     78\,459 & $+7.6\,\%$   &    183 \\
basicmath     &     962\,892 &  1\,172\,249 & $+21.7\,\%$  &  8\,500 \\
bitcount      &   2\,599\,349 & 3\,578\,829 & $+37.7\,\%$  & 67\,501 \\
blowfish      &   1\,463\,062 & 1\,509\,120 & $+3.1\,\%$   &  3\,005 \\
crc           &      11\,620 &     12\,469 & $+7.3\,\%$   &     13 \\
dijkstra      &   2\,151\,579 & 2\,150\,657 & $-0.0\,\%$   &      3 \\
limits        &       1\,977 &      4\,502 & $+127.7\,\%$ &    125 \\
qsort         &   1\,044\,383 & 1\,161\,010 & $+11.2\,\%$  &  7\,314 \\
randmath      &     149\,668 &    150\,135 & $+0.3\,\%$   &     11 \\
rc4           &      65\,128 &     67\,373 & $+3.4\,\%$   &      2 \\
rsa           &      48\,002 &     49\,058 & $+2.2\,\%$   &     51 \\
sha           &   1\,506\,595 & 1\,498\,536 & $-0.5\,\%$   &  1\,214 \\
\midrule
Geo.\ mean (all 14) & & & $+12.9\,\%$ & \\
Geo.\ mean (w/o limits\textsuperscript{a}) & & & $+7.0\,\%$ & \\
\bottomrule
\end{tabular}
\label{tab:mibench}
\smallskip
\raggedright\footnotesize
\textsuperscript{a}\,\emph{limits} is a 2k-cycle microbenchmark whose
runtime is entirely function-call overhead
\end{table}

\begin{figure}[h]
    \centering
    \includegraphics[scale=0.45]{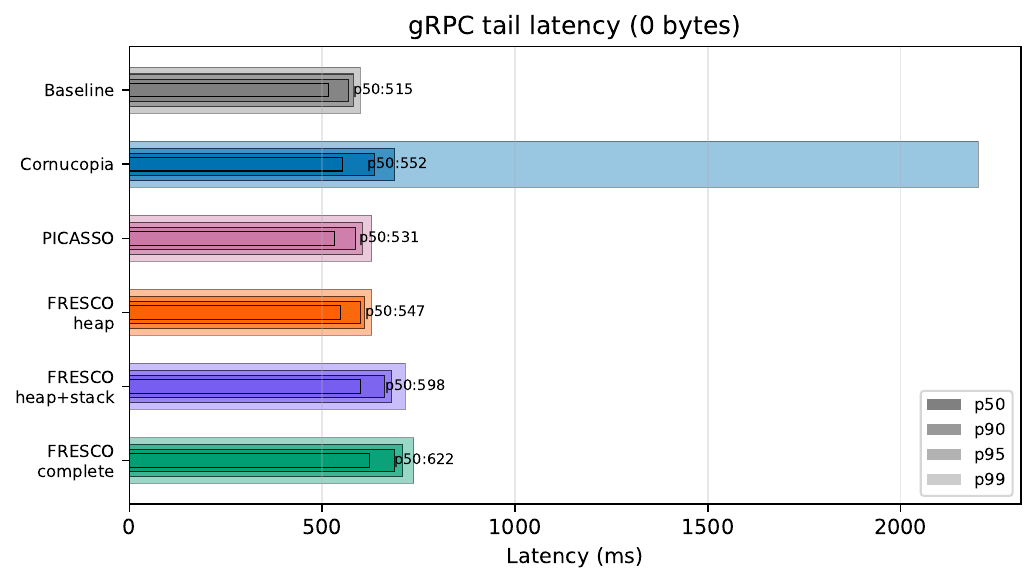}
    \caption{gRPC QPS latency percentile.}\label{fig:cc-qps}
\end{figure}
\begin{table*}
\centering
\small
\caption{Code size (\texttt{.text} bytes) of colored SPEC CPU2006 binaries}\label{tab:spec-code-size}
\begin{tabular}{lrrrr}
\toprule
Benchmark & Baseline & Full coloring & Color Saver & FRESCO complete \\
\midrule
401.bzip2 & 59512 & 65650 ($+10.3\,\%$) & 59884 ($+0.6\,\%$) & 60206 ($+1.2\,\%$) \\
445.gobmk & 670248 & 905938 ($+35.2\,\%$) & 677424 ($+1.1\,\%$) & 678574 ($+1.2\,\%$) \\
456.hmmer & 229232 & 274766 ($+19.9\,\%$) & 231228 ($+0.9\,\%$) & 232518 ($+1.4\,\%$) \\
458.sjeng & 105664 & 118490 ($+12.1\,\%$) & 107604 ($+1.8\,\%$) & 107606 ($+1.8\,\%$) \\
462.libquantum & 27436 & 36818 ($+34.2\,\%$) & 27436 ($+0.0\,\%$) & 27564 ($+0.5\,\%$) \\
464.h264ref & 594016 & 644514 ($+8.5\,\%$) & 595228 ($+0.2\,\%$) & 595982 ($+0.3\,\%$) \\
471.omnetpp & 413616 & 601438 ($+45.4\,\%$) & 422036 ($+2.0\,\%$) & 429004 ($+3.7\,\%$) \\
483.xalancbmk & 2314936 & 3571506 ($+54.3\,\%$) & 2426824 ($+4.8\,\%$) & 2456150 ($+6.1\,\%$) \\
\midrule
Geo.\ mean & & ($+26.5\,\%$) & ($+1.4\,\%$) & ($+2.0\,\%$) \\
\bottomrule
\end{tabular}
\end{table*}

\begin{figure*}[t]
    \centering
    \includegraphics[width=\linewidth]{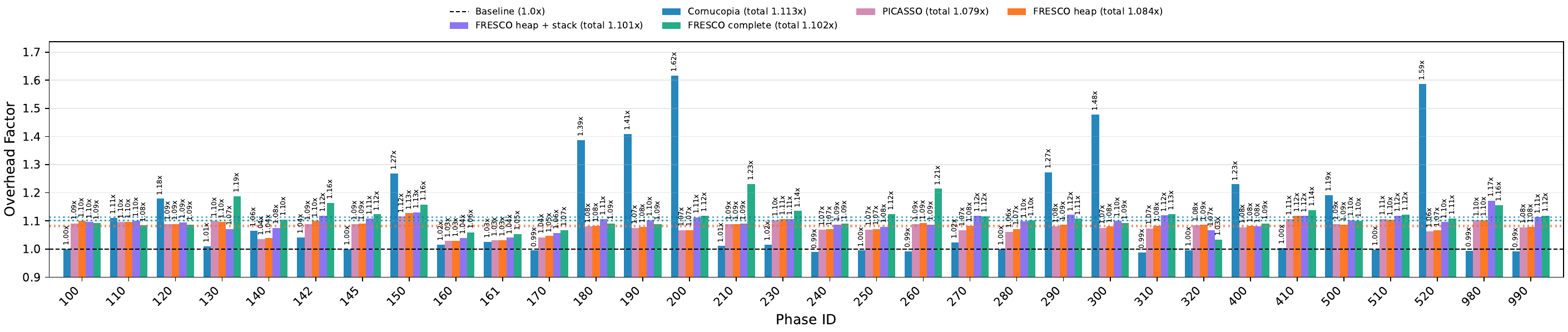}
    \caption{SQLite Speedtest1 Benchmark Results}\label{fig:cc-sqlite-phase}
\end{figure*}

\newpage
\onecolumn
\begin{sidewaystable}
\centering
\caption{SQLite Speedtest1 phase descriptions and overhead compared to baseline.}\label{tab:sqlitephase}
\label{tab:sqlite-fresco}
\resizebox{\textwidth}{!}{
\begin{tabular}{@{}llrrrrrrrrrr@{}}
\toprule
ID & Benchmark & \multicolumn{2}{c}{Cornucopia} & \multicolumn{2}{c}{PICASSO} & \multicolumn{2}{c}{FRESCO heap} & \multicolumn{2}{c}{FRESCO heap + stack} & \multicolumn{2}{c}{FRESCO complete} \\
\cmidrule(lr){3-4} \cmidrule(lr){5-6} \cmidrule(lr){7-8} \cmidrule(lr){9-10} \cmidrule(lr){11-12}
 & & Time (s) & Overhead & Time (s) & Overhead & Time (s) & Overhead & Time (s) & Overhead & Time (s) & Overhead \\
\midrule
100 & 50000 INSERTs into table with no index & 49.06 & 1.00$\times$ & 53.48 & 1.09$\times$ & 53.93 & 1.10$\times$ & 53.75 & 1.10$\times$ & 53.54 & 1.09$\times$ \\
110 & 50000 ordered INSERTS with one index/PK & 92.61 & 1.11$\times$ & 91.37 & 1.10$\times$ & 91.35 & 1.10$\times$ & 91.78 & 1.10$\times$ & 90.36 & 1.08$\times$ \\
120 & 50000 unordered INSERTS with one index/PK & 112.15 & 1.18$\times$ & 103.55 & 1.09$\times$ & 103.63 & 1.09$\times$ & 104.15 & 1.09$\times$ & 103.39 & 1.09$\times$ \\
130 & 25 SELECTS, numeric BETWEEN, unindexed & 96.00 & 1.01$\times$ & 104.66 & 1.10$\times$ & 104.12 & 1.10$\times$ & 101.75 & 1.07$\times$ & 112.82 & 1.19$\times$ \\
140 & 10 SELECTS, LIKE, unindexed & 129.20 & 1.06$\times$ & 125.58 & 1.04$\times$ & 126.19 & 1.04$\times$ & 130.49 & 1.08$\times$ & 133.97 & 1.10$\times$ \\
142 & 10 SELECTS w/ORDER BY, unindexed & 198.22 & 1.04$\times$ & 207.27 & 1.09$\times$ & 209.31 & 1.10$\times$ & 212.81 & 1.12$\times$ & 221.51 & 1.16$\times$ \\
145 & 10 SELECTS w/ORDER BY and LIMIT, unindexed & 151.94 & 1.00$\times$ & 165.48 & 1.09$\times$ & 165.59 & 1.09$\times$ & 168.32 & 1.11$\times$ & 170.67 & 1.12$\times$ \\
150 & CREATE INDEX five times & 163.38 & 1.27$\times$ & 143.72 & 1.12$\times$ & 145.22 & 1.13$\times$ & 145.57 & 1.13$\times$ & 149.16 & 1.16$\times$ \\
160 & 10000 SELECTS, numeric BETWEEN, indexed & 87.46 & 1.02$\times$ & 88.59 & 1.03$\times$ & 88.65 & 1.03$\times$ & 89.57 & 1.04$\times$ & 91.17 & 1.06$\times$ \\
161 & 10000 SELECTS, numeric BETWEEN, PK & 89.23 & 1.03$\times$ & 89.71 & 1.03$\times$ & 89.76 & 1.03$\times$ & 90.53 & 1.04$\times$ & 91.63 & 1.05$\times$ \\
170 & 10000 SELECTS, text BETWEEN, indexed & 167.54 & 0.99$\times$ & 175.43 & 1.04$\times$ & 176.21 & 1.05$\times$ & 178.14 & 1.06$\times$ & 179.66 & 1.07$\times$ \\
180 & 50000 INSERTS with three indexes & 200.89 & 1.39$\times$ & 156.29 & 1.08$\times$ & 156.64 & 1.08$\times$ & 160.12 & 1.11$\times$ & 157.91 & 1.09$\times$ \\
190 & DELETE and REFILL one table & 207.29 & 1.41$\times$ & 158.07 & 1.07$\times$ & 158.67 & 1.08$\times$ & 162.27 & 1.10$\times$ & 160.13 & 1.09$\times$ \\
200 & VACUUM & 182.75 & 1.62$\times$ & 120.56 & 1.07$\times$ & 120.67 & 1.07$\times$ & 125.68 & 1.11$\times$ & 126.48 & 1.12$\times$ \\
210 & ALTER TABLE ADD COLUMN, and query & 4.33 & 1.01$\times$ & 4.67 & 1.09$\times$ & 4.67 & 1.09$\times$ & 4.67 & 1.09$\times$ & 5.28 & 1.23$\times$ \\
230 & 10000 UPDATES, numeric BETWEEN, indexed & 80.19 & 1.02$\times$ & 86.99 & 1.10$\times$ & 87.34 & 1.11$\times$ & 87.29 & 1.11$\times$ & 89.59 & 1.14$\times$ \\
240 & 50000 UPDATES of individual rows & 111.52 & 0.99$\times$ & 120.51 & 1.07$\times$ & 120.63 & 1.07$\times$ & 122.40 & 1.09$\times$ & 122.91 & 1.09$\times$ \\
250 & One big UPDATE of the whole 50000-row table & 19.02 & 1.00$\times$ & 20.43 & 1.07$\times$ & 20.48 & 1.07$\times$ & 20.61 & 1.08$\times$ & 21.48 & 1.12$\times$ \\
260 & Query added column after filling & 4.57 & 0.99$\times$ & 5.02 & 1.09$\times$ & 5.03 & 1.09$\times$ & 5.01 & 1.09$\times$ & 5.60 & 1.21$\times$ \\
270 & 10000 DELETEs, numeric BETWEEN, indexed & 180.69 & 1.02$\times$ & 188.16 & 1.07$\times$ & 191.14 & 1.08$\times$ & 197.26 & 1.12$\times$ & 196.93 & 1.12$\times$ \\
280 & 50000 DELETEs of individual rows & 128.31 & 1.00$\times$ & 136.09 & 1.06$\times$ & 137.36 & 1.07$\times$ & 140.78 & 1.10$\times$ & 141.37 & 1.10$\times$ \\
290 & Refill two 50000-row tables using REPLACE & 399.53 & 1.27$\times$ & 338.96 & 1.08$\times$ & 341.18 & 1.09$\times$ & 351.96 & 1.12$\times$ & 347.68 & 1.11$\times$ \\
300 & Refill a 50000-row table using (b\&1)==(a\&1) & 219.51 & 1.48$\times$ & 159.50 & 1.07$\times$ & 160.30 & 1.08$\times$ & 163.37 & 1.10$\times$ & 162.04 & 1.09$\times$ \\
310 & 10000 four-ways joins & 319.78 & 0.99$\times$ & 347.18 & 1.07$\times$ & 350.50 & 1.08$\times$ & 362.28 & 1.12$\times$ & 364.09 & 1.13$\times$ \\
320 & subquery in result set & 649.39 & 1.00$\times$ & 707.26 & 1.08$\times$ & 709.12 & 1.09$\times$ & 695.58 & 1.07$\times$ & 673.62 & 1.03$\times$ \\
400 & 70000 REPLACE ops on an IPK & 126.69 & 1.23$\times$ & 110.89 & 1.08$\times$ & 111.43 & 1.08$\times$ & 111.26 & 1.08$\times$ & 112.30 & 1.09$\times$ \\
410 & 70000 SELECTS on an IPK & 71.51 & 1.00$\times$ & 78.81 & 1.11$\times$ & 79.61 & 1.12$\times$ & 79.58 & 1.12$\times$ & 81.05 & 1.14$\times$ \\
500 & 70000 REPLACE on TEXT PK & 130.57 & 1.19$\times$ & 119.23 & 1.09$\times$ & 118.94 & 1.09$\times$ & 120.84 & 1.10$\times$ & 120.48 & 1.10$\times$ \\
510 & 70000 SELECTS on a TEXT PK & 107.79 & 1.00$\times$ & 119.61 & 1.11$\times$ & 119.28 & 1.10$\times$ & 120.77 & 1.12$\times$ & 121.21 & 1.12$\times$ \\
520 & 70000 SELECT DISTINCT & 97.70 & 1.59$\times$ & 65.48 & 1.06$\times$ & 65.63 & 1.07$\times$ & 67.45 & 1.10$\times$ & 68.24 & 1.11$\times$ \\
980 & PRAGMA integrity\_check & 289.91 & 0.99$\times$ & 321.43 & 1.10$\times$ & 321.64 & 1.10$\times$ & 341.81 & 1.17$\times$ & 337.10 & 1.16$\times$ \\
990 & ANALYZE & 41.07 & 0.99$\times$ & 44.59 & 1.08$\times$ & 44.63 & 1.08$\times$ & 46.09 & 1.11$\times$ & 46.27 & 1.12$\times$ \\
\midrule
\textbf{Total} & (baseline 4410.23 s) & 4909.81 & 1.11$\times$ & 4758.56 & 1.08$\times$ & 4778.84 & 1.08$\times$ & 4853.95 & 1.10$\times$ & 4859.63 & 1.10$\times$ \\
\multicolumn{2}{l}{Cycles (baseline 110,297,124,849)} & 122,791,628,666 & 1.11$\times$ & 119,008,630,444 & 1.08$\times$ & 119,515,005,225 & 1.08$\times$ & 121,392,070,775 & 1.10$\times$ & 121,534,708,445 & 1.10$\times$ \\
\multicolumn{2}{l}{Instructions (baseline 61,758,979,977)} & 75,327,090,954 & 1.22$\times$ & 62,376,697,908 & 1.01$\times$ & 62,061,319,862 & 1.00$\times$ & 62,674,079,628 & 1.01$\times$ & 64,198,625,846 & 1.04$\times$ \\
\multicolumn{2}{l}{Memory (RSS) (baseline 14,304)} & 28,484 & 1.99$\times$ & 14,400 & 1.01$\times$ & 24,544 & 1.72$\times$ & 27,264 & 1.91$\times$ & 27,284 & 1.91$\times$ \\
\bottomrule
\end{tabular}}
\end{sidewaystable}

\end{document}